\pdfoutput=1

\documentclass[11pt, a4paper]{article}
\usepackage{jcappub}
\usepackage{amsmath}
\usepackage{graphicx,epsf,hyperref}
\usepackage{color}
\usepackage{xcolor}

\def\be{\begin{equation}}
\def\ee{\end{equation}}
\def\bea{\begin{eqnarray}}
\def\eea{\end{eqnarray}}
\def\lb{\left(}
\def\rb{\right)}
\def\lbs{\left[}
\def\rbs{\right]}

\def\p{\partial}
\def\n{\nabla}

\def\L{{\pounds}}

\def\T{\mathcal{T}}
\def\Hh{\mathcal{H}}

\newcommand\ul[1]{{\underline{#1}}}
\def\cs2{c_{\rm{s}}^2}
\def\U0{{\bar U_0}}
\def\wt{\widetilde}
\def\12{\frac{1}{2}}

\def\U{{\Upsilon}}

\title{Isocurvature initial conditions for second order Boltzmann solvers}
\author{Pedro Carrilho}
\author{and Karim A. Malik}

\affiliation{Astronomy Unit, School of Physics and Astronomy, Queen Mary University of London, Mile End Road, London, E1 4NS, UK}

\emailAdd{p.gregoriocarrilho@qmul.ac.uk}
\emailAdd{k.malik@qmul.ac.uk}

\abstract{We study how to set the initial evolution of general cosmological fluctuations at second order, after neutrino decoupling. We compute approximate initial solutions for the transfer functions of all the relevant cosmological variables sourced by quadratic combinations of adiabatic and isocurvature modes. We perform these calculations in synchronous gauge, assuming a Universe described by the $\Lambda$CDM model and composed of neutrinos, photons, baryons and dark matter. We highlight the importance of mixed modes, which are sourced by two different isocurvature or adiabatic modes and do not exist at the linear level. In particular, we investigate the so-called compensated isocurvature mode and find non-trivial initial evolution when it is mixed with the adiabatic mode, in contrast to the result at linear order and even at second order for the unmixed mode. Non-trivial evolution also arises when this compensated isocurvature is mixed with the neutrino density isocurvature mode. Regarding the neutrino velocity isocurvature mode, we show it unavoidably generates non-regular (decaying) modes at second order. Our results can be applied to second order Boltzmann solvers to calculate the effects of isocurvatures on non-linear observables.}

\begin{document}
\maketitle

\section{Introduction}

The importance of understanding the early Universe cannot be overstated. It would not only provide us with some insight into the complete history of the Universe, but would also allow one to probe new physics at energy scales beyond the capabilities of any other exploration method. However, the fact that the early Universe is not directly observable complicates its understanding considerably, since the only properties that can be accurately probed are the initial conditions for the stage in which the laws of physics are known. It is those initial conditions for the radiation dominated era that contain information about the state and evolution of the early Universe and it is through the measurement of their statistical properties that one can distinguish different models of the earliest epochs.

Observations of CMB anisotropies measured by the Planck satellite~\cite{Ade:2015lrj}, as well as measurements of the Large Scale Structure on complementary scales~\cite{Tegmark:2003uf,Cole:2005sx}, indicate the initial conditions to be adiabatic up to the uncertainty of the experiments~\cite{Beltran:2004uv}. The measured properties of this adiabatic mode have shown it to have a nearly scale invariant power spectrum and undetectably small non-Gaussianity~\cite{Ade:2015ava}. This has provided strong support for inflation, which generally predicts such adiabatic initial conditions to have nearly scale invariant properties.

However, many models of inflation with multiple fields, such as the curvaton model~\cite{Lyth:2001nq,Lyth:2002my,Gordon:2002gv}, also predict the existence of non-adiabatic modes~\cite{Linde:1984ti,Langlois:1999dw}, which, if measured in the future, would be essential in distinguishing different inflationary scenarios. Furthermore, there are certain non-adiabatic modes that can escape detection by conventional means. These are the so-called compensated isocurvature modes \cite{Grin:2011tf,He:2015msa,Heinrich:2016gqe,Valiviita:2017fbx}, which evade most of the constraints at the linear level by not producing an overall matter isocurvature mode. They are also conjectured to contribute to the flattening of the peaks of the CMB angular power spectrum for short scales, being degenerate with lensing~\cite{Heinrich:2016gqe,Valiviita:2017fbx}. Non-Gaussianity has remained undetected so far, and it is conceivable that it exists at similar levels for both adiabatic and non-adiabatic modes~\cite{Linde:1996gt,Langlois:2012tm,Hikage:2012tf}. In summary, non-adiabatic modes are still observationally relevant and their detection could open new windows into the physics of the early Universe.

Linear isocurvature modes have been the subject of many studies in the literature due to their potential to reveal more about the early universe~\cite{Suto:1984aa,Kodama:1986fg,Kodama:1986ud}. One of the first works on the general initial fluctuations was by Bucher, Moodley and Turok \cite{Bucher:1999re}, which defined the initial conditions after neutrino decoupling for a system with massless neutrinos, baryons, dark matter and photons. This was further extended to include more species such as primordial magnetic fields \cite{Finelli:2008xh, Paoletti:2008ck} and massive neutrinos \cite{Shaw:2009nf}. Among other results, these works defined the different isocurvature modes and calculated the initial evolution for a set of variables at first order in cosmological perturbation theory. This was then applied in a variety of Einstein-Boltzmann solvers \cite{Seljak:1996is,Lewis:1999bs,Lesgourgues:2011re} and later to the WMAP and Planck data to constrain isocurvature modes. 

As observational efforts move towards increasing precision, new observables will become available, whose predictions require calculations at non-linear orders in perturbation theory. Examples include the intrinsic bispectrum \cite{Pitrou:2010sn,Pettinari:2013he,Pettinari:2014vja}, magnetic field generation during the pre-recombination era \cite{Fenu:2010kh,Maeda:2011uq,Nalson:2013jya,Fidler:2015kkt} and vorticity production \cite{Christopherson:2009bt,Christopherson:2010ek,Christopherson:2010dw,Brown:2011dn}. In order to compute some of these observables, the initial evolution at second order for adiabatic modes has been calculated \cite{Pitrou:2010sn,Pettinari:2013he,Pettinari:2014vja}. However, these observables could also receive a sizeable contribution from isocurvature modes, in particular if the isocurvature spectrum is strongly blue or has features~\cite{Carrilho:2016con}. This is relevant also because of the mode mixing that occurs at non-linear orders in perturbation theory, which allows for an adiabatic mode to mix with an isocurvature one and generate contributions which are not present at linear order. These contributions should be the largest ones that involve isocurvatures at second order and should be the easiest to constrain or detect. These are the reasons that motivate us to investigate isocurvature modes at second order in a systematic way.
\\\\
In this paper, we calculate the initial evolution of cosmological fluctuations at second order in the presence of isocurvature modes. These calculations are essential for initializing Boltzmann codes at second order~\cite{Pettinari:2014vja} and thus to calculate observables with the required accuracy for comparing with experiment. We begin in Section \ref{defs} with the definitions and conventions we use in the remainder of the paper. In Section \ref{diffsys}, we describe the general differential system under study and how to split its perturbative solutions into different modes. After that, we introduce a clear definition of the isocurvature basis in Section \ref{isocurv} as used in previous literature and present our results for the initial time evolution in synchronous gauge in Section \ref{inievo}. We then discuss our results and conclude in Section \ref{conclusion}. We also consider gauge transformations and calculate the specific transformation of our results into Poisson gauge in appendix \ref{gaugetr} and discuss the Liouville equation at second order in appendix \ref{boltzapp}.

\section{Cosmological perturbation theory}\label{defs}

In this first section, we introduce our notation and conventions for the metric
and stress-energy tensors that will be used in the rest of the paper. We follow most of the notation
of Ref.~\cite{Malik:2008im}, except for the coefficient of the tensor mode in the definition of the metric, as is clear in Eq.~\eqref{gij} below.\\
All quantities, $T$, are expanded as
\bea
T=T^{(0)}+\delta T^{(1)}+\frac12\delta T^{(2)}+...\,,
\eea
with the superscript denoting the order in perturbation theory. We drop the 0th order superscript for simplicity of presentation.

We assume an FLRW background spacetime with zero
spatial curvature throughout the paper, and use conformal time. Greek indices,
$\mu,\nu,\lambda$, etc., range from $0$ to $3$, while lower case Latin
indices, $i,j,k$, denote spatial indices ranging from $1$ to $3$.
\\\\
The metric tensor can be split in several different ways, which vary in the way the spatial part of the
metric is arranged \cite{Carrilho:2015cma}. The version that we will use in the
calculations below takes the following form, at all orders
\begin{align}
g_{00}=&-a^2\lb 1+2\phi\rb\,,\label{metric1}\\
g_{i0}=&a^2\lb B_{,i}-S_i\rb\,,\\
g_{ij}=&a^2\lb\delta_{ij}+2C_{ij}\rb\label{metric3}\,,
\end{align}
in which $\phi$ is the perturbation to the lapse, $B$ and $S_i$ are,
respectively, the scalar and vector parts of the shift and $C_{ij}$ is
the perturbation to the spatial part of the metric, which is further split as
\begin{equation}
\label{gij}
C_{ij}=-\psi\delta_{ij}+E_{,ij}+F_{(i,j)}+h_{ij}\,,
\end{equation}
in which $\psi$ is the curvature perturbation, $E$ and $F_i$ are, respectively, the scalar and vector
potentials of the traceless part of the spatial metric and $h_{ij}$ is the tensor potential, usually understood to
represent gravitational waves.

As for the total stress-energy tensor, we choose the so-called energy
frame to represent it \cite{EMM,HwangNoh}\footnote{This frame is defined by the
condition that the 4-velocity, $u^\mu$, is an eigenvector of
$T_{\mu\nu}$, with eigenvalue $\rho$. This is equivalent to setting
the energy flux, $q^\mu$, to zero.}, and include anisotropic stress:
\be
T_{\mu\nu}=\lb P+\rho\rb u_\mu u_\nu+P g_{\mu\nu}+\pi_{\mu\nu}\,,
\ee
with
\be
\label{constrPi}
\pi^\mu_{\ \mu}=0\,,\ \  \text{  and }\pi_{\mu\nu}u^\mu=0\,. 
\ee
The perturbative expansion is the standard one, as given
in Ref.~\cite{Malik:2008im} and in Ref.~\cite{Carrilho:2015cma} for $\pi_{\mu\nu}$. We reproduce here the scalar-vector-tensor (SVT) splitting of
$\pi_{ij}^{(1)}$,
\be
\pi_{ij}^{(1)}=a^2\lbs\Pi_{ij}^{(1)}+\Pi_{(i,j)}^{(1)}+\Pi_{,ij}^{(1)}-\frac{1}{3}\delta_{ij}\n^2\Pi^{(1)}\rbs\,,
\ee
where the quantities $\Pi$, $\Pi_i$ and $\Pi_{ij}$ are, respectively, the scalar, vector and tensor parts of the linear piece of the anisotropic stress tensor. We also define the variable $\sigma$, at all orders,
\be
\sigma^{(i)}=-\frac1{2\rho}\n^2\Pi^{(i)}\,,
\ee
which is more appropriate here as it is more directly linked to the conventions used in the literature and, as we shall see below, has growing mode solutions.

The stage of the evolution of the Universe we study in this work is the radiation dominated epoch at the time following neutrino decoupling. At this stage, (in the $\Lambda$CDM model) there are four matter species that are present in the Universe, namely, neutrinos ($\nu$), photons ($\gamma$), baryons ($b$) and cold dark matter ($c$). We construct the total stress-energy tensor by adding those of each species, labelled by the index $s$,
\be
T^{\alpha\beta}=\sum_s T_s^{\alpha\beta}\,,
\ee
which are given by
\begin{align}
&T_c^{\alpha\beta}=\rho_c u_c^\alpha u_c^\beta\,,\\
&T_b^{\alpha\beta}=\rho_b u_b^\alpha u_b^\beta\,,\\
&T_\gamma^{\alpha\beta}=\frac43\rho_\gamma u_\gamma^\alpha u_\gamma^\beta+\frac13 \rho_\gamma g^{\alpha\beta}\,,\\
&T_\nu^{\alpha\beta}=\frac43\rho_\nu u_\nu^\alpha u_\nu^\beta+\frac13 \rho_\gamma g^{\alpha\beta}+\pi_\nu^{\alpha\beta}\,.
\end{align}
It is clear from these expressions that only neutrinos have anisotropic stress, as it is assumed that photons are tightly coupled with baryons at this time and dark matter is too cold to have appreciable anisotropic stress. We thus set it to zero. Note as well that all species have been written in their specific energy frames given by each 4-velocity vector $u_s^\alpha$. This implies that the calculation of the total fluid quantities, such as the total energy density, is not a simple sum of those variables defined in each frame. We perform this calculation by projecting the stress-energy tensors of each species into a global energy frame, labelled by the 4-velocity vector $u^\mu$. After this change of frame, we find the total energy density, pressure and anisotropic stress are given by
\begin{align}
\label{rhotot}
\rho=&\gamma_c^2\rho_c+\gamma_b^2\rho_b+\frac13\left(4\gamma_\gamma^2-1\right)\rho_\gamma+\frac13\left(4\gamma_\nu^2-1\right)\rho_\nu+\pi_\nu^{\alpha\beta}u_\alpha u_\beta \,,\\
P=&\frac13(\gamma_c^2-1)\rho_c+\frac13(\gamma_b^2-1)\rho_b+\frac19\left(4\gamma_\gamma^2-1\right)\rho_\gamma+\frac19\left(4\gamma_\nu^2-1\right)\rho_\nu +\frac13\pi_\nu^{\alpha\beta}u_\alpha u_\beta\,,\\
\pi^{\alpha\beta}=&\pi_\nu^{\alpha\beta}-\frac13(g^{\alpha\beta}+4u^{\alpha}u^{\beta})\pi_\nu^{\mu\lambda}u_\mu u_\lambda\nonumber\\
&+\sum_s(1+w_s)\left(\frac{1}{3}(1-\gamma_s^2)g^{\alpha\beta}+\frac{1}3(1-4\gamma_s^2)u^{\alpha}u^{\beta}+u_s^{\alpha}u_s^{\beta}\right)\rho_s\,,\label{eqpitot}
\end{align}
while the 4-velocity of the energy frame can be related to that of each fluid by solving the following equation for $u^\alpha$:
\be
\sum_s(1+w_s)\rho_s \gamma_s (u_s^\alpha-\gamma_s u^\alpha)-\pi_\nu^{\alpha\beta}u_\beta-\pi_\nu^{\mu\beta}u_\mu u_\beta u^\alpha=0\,,
\ee
which is obtained from the energy frame condition, i.e. by setting the momentum transfer vector $q^\alpha$ to zero. In the absence of neutrino anisotropic stress, one would find the following solution for $u^\alpha$:
\be
u^\alpha=\frac{\sum_s(1+w_s)\rho_s \gamma_s u_s^\alpha}{\sum_s(1+w_s)\rho_s \gamma_s^2}\,.
\ee
This result is still correct at first order, but is not sufficient at second order. In all expressions above, $w_s=P_s/\rho_s$ is the equation of state parameter and $\gamma_s$ is the Lorentz factor for changing between the energy frame and each species', which is given by 
\be
\gamma_s=-u_s^\lambda u_\lambda.
\ee
All these equations are fully covariant and are therefore valid at all orders in perturbation theory. In the following we will use them at second order.

\subsection{Evolution equations}

To describe the evolution of this system we assume Einstein gravity,
\be
G^{\alpha\beta}=8\pi G T^{\alpha\beta}\,,
\ee
and describe the evolution of each fluid by:
\begin{align}
&\n_\beta T_\gamma^{\alpha\beta}=C^\alpha_{\gamma b}\,,\\
&\n_\beta T_\nu^{\alpha\beta}=0\,,\\
&\n_\beta T_b^{\alpha\beta}=-C^\alpha_{\gamma b}\,,\\
&\n_\beta T_c^{\alpha\beta}=0\,.
\end{align}
where we have included the interaction of photons with baryons, represented by $C^\alpha_{\gamma b}$. However, we will assume the tight coupling approximation (TCA) is valid, which means that the velocity of the photons and baryons is equal. For the case of the neutrinos, we also introduce an equation for the anisotropic stress, which is derived from the Liouville equation. We shall write these equations below in their perturbed versions. We write only the second order equations as the first order ones can be obtained straightforwardly by setting all the non-linear terms to zero. Furthermore, we write all equations in the synchronous gauge, since that is the gauge we will use in most of the paper. We also include only scalars as we are only studying second-order scalar modes sourced by first-order scalars. We leave the study of vector and tensor modes for future work.

We begin by writing the field equations for the two scalar potentials available in synchronous gauge. The only ones we require are the constraint equations, given by
\begin{align}
&\n^2\psi+\Hh\n^2E'-3\Hh\psi'-\frac32\Hh^2\sum_s{\Omega_s\delta_{s}}=6\Hh\psi\psi'-\frac32(\psi')^2-4\psi\n^2\psi-\frac32\psi_{,i}\psi^{,i}\nonumber\\
&-2(\psi\n^2E)'+\psi'\n^2E'+\n^2E_{,i}\psi^{,i}+\n^2E\n^2\psi+\psi_{,ij}E^{,ij}-\frac14\n^2E'\n^2E'\label{EEq00}\\
&+\frac14\n^2E_{,i}\n^2E^{,i}+2\Hh E'_{,ij}E^{,ij}+\frac14E'_{,ij}E^{\prime,ij}-\frac14E_{,ijk}E^{,ijk}+\frac32\Hh^2\sum_s{(1+w_s)\Omega_s v_{s,i}v_{s}^{,i}}\,,\nonumber
\end{align}
and
\begin{align} 
&\psi'-\frac32\Hh^2\sum_s{(1+w_s)\Omega_s v_{s}}=-2(\psi\n^2\psi)'-4\psi'_{,i}\psi^{,i}+\n^2E'_{,i}\psi^{,i}+\frac12\n^2E'\n^2\psi\nonumber\\
&+\n^2E\n^2\psi'+\psi'_{,ij}E^{,ij}+\frac12\psi_{,ij}E^{\prime\,,ij}+\frac12\n^2E'_{,i}\n^2E^{,i}-\frac12E'_{,ijk}E^{,ijk}\label{EEq0i}\\
&-\frac34\Hh^2\sum_s{\Omega_s(1+w_s)\left[2\left((\delta_{s}-2\psi)v_{s}^{,i}\right)_{,i}+(v_{s,i}E^{,ij})_{,j}\right]}\nonumber\\
&-\Omega_\nu\Hh^2\left[(\sigma_{\nu}v_{\nu}^{,i})_{,i}-3(\n^{-2}\sigma_{\nu}^{,ij} v_{\nu,i})_{,j}\right]\,,\nonumber
\end{align}
in which $\Omega_s=8\pi G \rho_s/3H^2$ is the standard density parameter for each species, $\delta_s$ in the density contrast for each species, defined by $\delta_s=\delta\rho_s/\rho_s$, $v_s$ is the corresponding velocity fluctuation and $\sigma_\nu$ represents the scalar part of the neutrino anisotropic stress. The energy conservation equations for the fluids are given by
\begin{align}
&\delta_{s}'-(1+w_s)\left(3\psi'-\n^2(E'+v_{s})\right)=2(1+w_s)\left(3\psi\psi'-(\psi\n^2E)'+E'_{,ij}E^{,ij}\right)\nonumber\\
&+\delta_{s}\delta_{s}'-(1+w_s)v_{s}^{,i}\left(2v'_{s,i}+\delta_{s,i}-3\psi_{,i}+\n^2E_{,i}+(1-3w_s)\Hh v_{s,i}\right)\label{Eqdelta}\\
&+\frac23 \delta_s^\nu \left[2\sigma_{\nu,i}v_{\nu}^{,i}-\sigma_{\nu}\n^2(E'+v_{\nu})+3\n^{-2}\sigma_{\nu,ij}(E'+v_{\nu})^{,ij}\right]\,,\nonumber
\end{align}
where we have assumed that each fluid has a constant equation of state and have aggregated all possible cases for the four species under study. $\delta_s^\nu$, appearing the last line of Eq~\eqref{Eqdelta}, is the Kronecker delta symbol and is unrelated to the density contrast. 

Concerning the momentum conservation equations, we only have to write them for the neutrinos and the photon-baryon plasma. This is due to having chosen the synchronous gauge, which allows one to set the cold dark matter velocity to zero to fix the residual gauge conditions. Furthermore, since we assume the TCA is valid, there is only one equation for the common velocity of photons and baryons, $v_{b\gamma}$. This equation is obtained by summing the two momentum conservation equations for baryons and photons and is given by
\begin{align}
&\n^2\left[(3\Omega_b+4\Omega_\gamma)v_{b\gamma}'+\Omega_\gamma\delta_{\gamma}+3\Omega_b\Hh v_{b\gamma}\right]=-4\Omega_\gamma\left(\delta_{\gamma}v_{b\gamma}^{\prime,i}\right)_{,i}-3\Omega_b\left(\delta_{b}v_{b\gamma}^{\prime,i}\right)_{,i}\nonumber\\
&+v_{b\gamma}^{,i}\left[\Omega_\gamma\left(4\psi_{,i}-\frac{20}{3}\n^2E'_{,i}-\frac{8}{3}\n^2v_{b\gamma,i}\right)+\Omega_b\left(6\psi_{,i}-6\n^2E'_{,i}-3\n^2v_{b\gamma,i}-3\Hh\delta_{b,i}\right)\right]\nonumber\\
&-2\Omega_\gamma\left(\psi\delta_{\gamma}^{,i}-E^{,ij}\delta_{\gamma,j}\right)_{,i}+\n^2v_{b\gamma}\left[\Omega_\gamma\left(4\psi'+\frac43\n^2E'+\frac43\n^2v_{b\gamma}\right)+\Omega_b\left(6\psi'-\frac12\delta_{b}\right)\right]\nonumber\\
&-v_{b\gamma}^{,ij}\left(4\Omega_\gamma+3\Omega_b\right)\left(2E_{,ij}'+v_{b\gamma,ij}\right)\,,
\end{align}
while the one for neutrinos is given by
\begin{align}
&\n^2\left[v_{\nu}'+\frac14\delta_{\nu}+\sigma_{\nu}\right]=\frac12\left(\delta_{\gamma}^{,i}(E_{,ij}-\psi\delta_{ij})-v_{b\gamma}^{,i}(4E_{,ij}+2v_{b\gamma,ij})\right)^{,j}\nonumber\\
&-\left((\delta_{\nu}'-5\psi'+\n^2E'+\n^2 v_{\nu})v_{\nu}^{,i}-\delta_{\nu}v_{\nu}^{\prime\,,i}\right)_{,i}\label{Eqvnu2}\\
&+\left(\psi\sigma_{\nu,i}+\frac12\psi_{,i}\sigma_{\nu}-\frac32\psi^{,j}\n^{-2}\sigma_{\nu,ij}-\frac12(\sigma_{\nu}v_{\nu,i}-3v_{\nu}^{,j}\n^{-2}\sigma_{\nu,ij})'\right)^{,i}\nonumber\\
&-\frac12\left(\frac23\n^2E\sigma_{\nu,i}+E_{,ij}\sigma_{\nu}^{,j}-\frac43\n^2E_{,i}\sigma_{\nu}+5E^{,jk}\n^{-2}\sigma_{\nu,ijk}+4\n^2E^{,j}\n^{-2}\sigma_{\nu,ij}\right)^{,i}\nonumber\,.
\end{align}
The equation for $\sigma_\nu$ is derived from the Liouville equation, as explained in Appendix \ref{boltzapp}, following most of the conventions of Ref.~\cite{Pettinari:2014vja}. In synchronous gauge, that equation is given by
\begin{align}
\label{Boltz}
&\Delta_{T\,ij}'+\left(\Delta_{T\,ijk,l}-\frac15\left(\frac23\delta_{ij}\delta_k^{r}-\delta_{kj}\delta_i^{r}-\delta_{ik}\delta_j^r\right)\Delta_{r,l}\right)(\delta^{kl}-C^{kl})-\Delta_T^{ijkl}E'_{,kl}\\
&-4\Delta_{T\,ij} \psi'-\frac{10}{21}\delta^{ij}\Delta_T^{kl}E'_{,kl}+\frac17\left(6\Delta_T^{ij}\n^2E'+5\Delta_T^{ki}E^{\prime\,,j}_{,k}+5\Delta_T^{kj}E^{\prime\,,i}_{,k}\right)+\frac{8}{15}\Delta_0E^{\prime\,,ij}\nonumber\\
&-\frac{8}{45}\Delta_0\delta^{ij}\n^2E'-\left(4\Delta_{T\,ij}^{\ \ k}+\frac15\left(\frac23\delta_{ij}\delta^{ks}-\delta^k_{j}\delta_i^{s}-\delta_{i}^{k}\delta_j^s\right)\Delta_s\right)\psi_{,k}\nonumber\\
&+\frac{8}{15}\left[C^{ij}-C_k^iC^{kj}-\frac13\delta^{ij}(C^k_k-C_{kl}C^{kl})\right]'=0\,,\nonumber
\end{align}
in which the $\Delta$ variables are perturbations to the different moments of the distribution function of neutrinos and are called here brightness tensors. Their expressions are given in detail in appendix \ref{boltzapp}. While the third and higher rank tensors are defined solely in terms of the integrals of the distribution function, the first three can be related to the stress-energy tensor as follows:
\begin{align}
\label{d0}
&\Delta_0=-\frac{\delta T^{\ 0}_{\nu\ 0}}{\rho_\nu}\,,\\
\label{d1}
&\Delta^{i}=-\frac{T^{\ j}_{\nu\ 0}}{\rho_\nu}(\delta^i_j+C^i_j)\,,\\
\label{d2}
&\Delta^{\ \  i}_{T\, j}=\frac{1}{\rho_\nu}\left(T^{\ k}_{\nu\ l}-\frac13\delta^k_{\ l}T^{\ r}_{\nu\ r}\right)\left(\delta_{\ j}^l\delta_{\ k}^i+\delta_{\ j}^l E^{,i}_{\ ,k}-\delta_{\ k}^i E^{,l}_{\ ,j}\right)\,,
\end{align}
in which $\rho_\nu$ is the background neutrino energy density. Because we are only dealing with scalar modes, we compute the scalar part of Eq.~\eqref{Boltz} by applying the differential operator $\partial^i\partial^j$. Due to its complexity, we refrain from showing the final evolution equation for $\sigma_\nu$ here. It can be calculated straightforwardly from the scalar equation by using the conversion from the scalar part of $\Delta^{\ \  i}_{T\, j}$ to $\sigma_\nu$ shown at the end of appendix \ref{boltzapp}.

This concludes the description of the evolution equations. In the next sections we will describe this differential system in general and provide details about its formal solution.

\section{Differential System}\label{diffsys}

It is straightforward to show, after applying a Fourier transform, that the differential system presented in the previous section can be described by the following generic equation at any specific non-background order:
\be
\label{diffsystem}
\mathcal{D}_\tau X=Q(\tau)\,,
\ee
in which $\mathcal{D}_\tau$ is a \textbf{\emph{linear}} differential operator, $X$ is a vector including all the variables to evolve and $Q(\tau)$ includes all the non-linear terms, which act as a source at orders higher than the first, while at the linear level we have $Q^{(1)}=0$, by definition. For example, at second order, the source term is a convolution of squares of the first order (or linear) solutions, 
\be
Q^{(2)}(\tau,k)\supset\int_q X^{(1)}(\vec q-\vec k)X^{(1)}(\vec q)\,,
\ee
in which we introduce the notation
\be
\int_q =\int\frac{\text{d}^3q}{(2\pi)^3}\,.
\ee

In order to solve such a system, one begins by solving the first order equations. Being linear, the solutions to those equations can be written as a sum of particular solutions, the number of which is the same as the dimension of the solution space, $D$. The solution can therefore be written as
\be
X^{(1)}(\tau,k)=\sum_{i=1}^D{\mathcal{T}_i(\tau,k) I_i^{(1)}(k)}\,,
\ee
in which $\T_i(\tau,k)$ are transfer functions and $I_i(k)$ represent the initial conditions of certain variables of interest. These variables will be called the defining variables of a mode, since they are non-zero only when a specific mode is present. Each of the $\T_i$ is a vector (just like $X$) while each of the $I_i$ is a scalar. The $I_i$ are usually random variables which encode all the statistical information of the initial conditions, and, given that the evolution of the transfer functions is classical, they will allow us to calculate the statistics of $X^{(1)}$ at any time. The fact that each of the $\T_i(\tau)$ is an independent solution of the differential system also means that we can separate the numerical solution of the equations mode by mode, solving each one separately and later calculating the required statistics by summing all the modes. This is especially useful, since it allows for a solution of the equations without the need to specify the amplitude of each initial condition, leaving those parameters to be constrained by experiment.

At second order, the general solution is
\be
X^{(2)}(\tau,k)=\sum_{i}{\mathcal{T}_i(\tau,k) I_i^{(2)}(k)}+\sum_{i,j}{\int_{k_1,k_2}\mathcal{T}^{(2)}_{ij}(\tau,k,k_1,k_2) I_i^{(1)}(k_1)I_j^{(1)}(k_2)}\,,
\ee
in which the first term is the homogeneous solution to Eq.~\eqref{diffsystem}, i.e. it is the same solution as the first order one, only with different coefficients $I_i^{(2)}$. Given that fact, the total solution, up to this order, can be written as
\begin{align}
X(\tau,k)&=X^{(1)}(\tau,k)+\frac12X^{(2)}(\tau,k)\\
&=\sum_{i}{\mathcal{T}_i(\tau,k) \lb I_i^{(1)}(k)+\frac12I_i^{(2)}(k)\rb}+\frac12\sum_{i,j}{\int_{k_1,k_2}\mathcal{T}^{(2)}_{ij}(\tau,k,k_1,k_2) I_i^{(1)}(k_1)I_j^{(1)}(k_2)}\,,\nonumber
\end{align}
which shows that one can absorb the term $I_i^{(2)}$ into the first order part $I_i^{(1)}$ or, equivalently, setting $I_i^{(2)}=0$. In this case the defining variables $I_i=I_i^{(1)}+\frac12I_i^{(2)}$ are set by the initial conditions of the full $X$ and not just its first order part. This is also more natural, as many times, the initial conditions will not be split into different orders, unless they have different properties, such as non-Gaussianity. An alternative scenario is to write $I_i^{(2)}$ as a sum of $I_i^{(1)} I_j^{(1)}$, effectively including it into the second term above. This is also equivalent to the previous case, because nothing constrains $\T^{(2)}_{ij}$ from including terms proportional to $\mathcal{T}_i$.

To numerically solve the differential system in question one may also separate the solution of the different transfer functions $\T^{(2)}_{ij}$, in order to find solutions which are valid for any values of the amplitude of the initial conditions. To see why this split can be performed, we begin by analysing the source $Q(\tau,k)$. It can also be written in terms of the defining variables as:
\be
Q^{(2)}(\tau,k)=\sum_{i,j}{\int_{k_1,k_2}\mathcal{S}_{ij}(\tau,k,k_1,k_2) I_i(k_1)I_j(k_2)}\,,
\ee
in which $\mathcal{S}_{ij}$ are the equivalent of transfer functions for the source terms $Q^{(2)}$. It can be shown, due to the linearity of the differential system, that there is a particular solution to the second order system which is a sum of the solutions of similar systems with the source $Q^{(2)}$ substituted for each of the terms in the sum above. Hence, to find the evolution of each $\T^{(2)}_{ij}$ one needs only to solve those similar systems in which only the $\{i,j\}$ defining variables are non-zero.

The question that we are concerned with in this paper is that of the initial evolution of $\T^{(2)}_{ij}$, to be used in setting up its numerical evolution. The aim is to find an approximation to the transfer functions that is valid when all Fourier modes of interest are still super-horizon during the radiation dominated Universe. In the following section, we precisely define the isocurvature basis. 

\section{Definition of isocurvature basis}\label{isocurv}

In the radiation dominated Universe and after neutrino decoupling at $z\sim10^9$, the species that are relevant are (nearly) massless neutrinos, the dark matter fluid and the tightly coupled baryon-photon plasma. In the case that those species can be represented by barotropic perfect fluids, one can show that the total number of evolving scalar degrees of freedom is 8. This is due to the fact that, for each fluid, the perturbed energy conservation equation and
the momentum conservation equation allow us to derive a second order
ODE (in $k$-space). In an appropriate gauge, such as flat gauge~\cite{Christopherson:2010ek}, one may use the Einstein constraint equations to
eliminate the metric potentials, and arrive at a system only in terms
of fluid quantities, such as energy densities, pressures, etc. To close the system, one uses the barotropic and perfect nature of the fluids to set the entropy and anisotropic stress fluctuations to zero. Finally, one specifies an equation of state, relating pressure and energy density, which results in a second order ODE for the density perturbation of each fluid. Thus, for each barotropic perfect fluid there are 2 independent modes, hence 8 in total\footnote{The situation is slightly different in synchronous gauge. In that case, one of the metric potentials cannot be completely eliminated from the final equations in terms of the density contrasts. Therefore an extra equation for that potential is required, which appears to increase the number of degrees of freedom to 9. This is a peculiarity of this gauge, for which the coordinate freedom has not been exhausted. The 9th mode is in fact a gauge mode, which is often eliminated by setting the initial velocity field of the dark matter fluid to zero.}. Naturally, there may be more modes, if, like the neutrinos, the fluids are not perfect. However, it is unlikely that those modes are present if the fluid has been tightly coupled in the past, as such a stage brings any anisotropic stress to negligible values. After decoupling, an anisotropic stress perturbation will be generated, but only after horizon re-entry.

However, as is well known in the literature \cite{Bucher:1999re}, only 5 of the 8 modes are growing modes in the standard case. This reduction from the total 8 degrees of freedom is due, firstly, to tight coupling, which forces the velocities of baryons and photons to be equal, or, in other words, constrains the mode generated by their difference to be a rapidly decaying mode. Two more modes are also decaying modes, and, in synchronous gauge, arise due to the presence of a non-zero total density contrast and total velocity, as can be seen by analysing the first order versions of Eqs.~\eqref{EEq00} and \eqref{EEq0i}. Setting those quantities to zero eliminates the corresponding decaying modes at first order. The five remaining independent modes are usually represented in the so-called isocurvature basis, in which one defines an adiabatic mode and 4 isocurvature modes: dark matter, baryon and neutrino density isocurvatures as well as the neutrino velocity isocurvature, which are labelled in accordance to the defining variable, $I_i$ that is non-zero in each mode. All observational evidence points towards the adiabatic mode being the dominant one and that is why it is used to define this basis. The other modes could possibly be split in different ways, but we stick here to the conventions of the literature, as this parametrisation is commonly used in observational studies.

At second order, an interesting issue arises. Looking again at Eqs~\eqref{EEq00} and \eqref{EEq0i}, we see that the terms proportional to $\Hh^2$ are responsible for generating decaying solutions, since $\Hh\approx \tau^{-1}$ during radiation domination. In order to make sure those terms are disabled, we actually require $\delta=v=0$ \emph{and} $v'=\delta'=0$ at the initial time. At first order, however, the condition on the derivatives is a consequence of the original condition, $\delta=v=0$, as can be shown by checking the total energy and momentum conservation equations:
\begin{align}
\delta'+(1+w)\left(\frac32\Hh\delta+\n^2v-\frac{1}{\Hh}\n^2\psi\right)=0\,,\\
(1+w)v'+(1+w)(1-3w)\Hh v+w\delta-\frac43\sigma=0\,,\label{totvpeq}
\end{align}
in which $w=P/\rho$ is the equation of state parameter for the total fluid. To show that these imply $v'=\delta'=0$ when $\delta=v=0$, we first note that $\Hh^{-1}\approx \tau$ and as a consequence the term with $\psi$ is negligible initially. The second and crucial step is noticing that $\sigma$ is initially zero at first order, because it is proportional to the neutrino anisotropic stress. At second order, this is no longer true, since the total anisotropic stress depends on the velocity fluctuations of each species, as can be shown from Eq.~\eqref{eqpitot}\footnote{Contributions from non-linear terms appearing in the second order version of Eq.~\eqref{totvpeq} are not important for this argument as they can be shown to be initially zero for all possible growing modes at first order.}. Therefore, the conditions required for non-decaying solutions are not satisfied at second order in all cases. In particular, we expect the neutrino velocity mode to have a decaying component at second order, since it is the only one for which the velocities of particular species are initially non-zero. For this reason, we choose not to perform any calculations at second order with the neutrino velocity mode. We now describe the standard way of performing the general decomposition, including the description of the neutrino velocity mode, for completeness.

We begin with the adiabatic mode. It is defined to be the mode whose initial conditions have vanishing entropy perturbations and vanishing velocity for all species. At first order, the gauge invariant relative entropy perturbation is given by (\cite{Malik:2002jb})
\be
S_{sr}=3 (\zeta_s-\zeta_r)\,,
\ee
in which $r$ and $s$ label the species in question and $\zeta_s$ is the partial curvature perturbation of species $s$, which is given by
\be
\label{zetai}
\zeta_{s}=-\psi+\frac{\delta_s}{3(1+w_s)}\,,
\ee
where we have assumed that energy transfer is negligible. In order to define any general mode one must give five initial conditions, as that is the number of growing modes present in the system. However, we wish to leave one of those initial conditions free --- the amplitude of the corresponding mode --- so that it may later be fixed by measurement. Thus, we only present four conditions for each mode. For the adiabatic one, the conditions are, in terms of the relative entropies:
\begin{gather}
\label{Adcond1}
S_{c\gamma}|_{\tau=0}=S_{\nu\gamma}|_{\tau=0}=S_{b\gamma}|_{\tau=0}=S_{c\nu}'|_{\tau=0}=0\,,
\end{gather}
In synchronous gauge, in which these conditions were originally defined, the adiabatic mode is given in terms of density contrasts and the neutrino velocity:
\begin{gather}
\label{Adcond2}
\delta_{c}|_{\tau=0}=\delta_{\nu}|_{\tau=0}=\delta_{b}|_{\tau=0}=v_\nu|_{\tau=0}=0\,.
\end{gather}
We can show that these conditions are equivalent to the ones for the entropies as $\delta_\gamma|_{\tau=0}=0$ due to the total density contrast being set to zero to avoid decaying modes. The defining variable in this case is $\psi|_{\tau=0}=-\zeta|_{\tau=0}$.

For the isocurvature modes, instead of the initial entropy being zero, these modes require the initial curvature perturbation, $\zeta$, to vanish. The different density isocurvature modes are then distinguished from each other by the fact that at least one of the density contrasts (or neutrino velocity) is initially non-zero. 

We summarize here all the conditions for the isocurvature modes at first order in perturbation theory, written in synchronous gauge:

\textbf{Baryon isocurvature:}
\begin{align}
\label{bicond}
&\delta_{c}|_{\tau=0}=\delta_{\nu}|_{\tau=0}=\psi|_{\tau=0}=v_{\nu}|_{\tau=0}=0\,,\\
&\text{Defining variable: }\delta_b.\nonumber
\end{align}

\textbf{Cold dark matter isocurvature:}
\begin{align}
\label{cdicond}
&\delta_{b}|_{\tau=0}=\delta_{\nu}|_{\tau=0}=\psi|_{\tau=0}=v_{\nu}|_{\tau=0}=0\,,\\
&\text{Defining variable: }\delta_c.\nonumber
\end{align}

\textbf{Neutrino Density Isocurvature:}
\label{nidcond}
\begin{align}
&\delta_{c}|_{\tau=0}=\delta_{b}|_{\tau=0}=\psi|_{\tau=0}=v_{\nu}|_{\tau=0}=0\,,\\
&\text{Defining variable: }\delta_\nu.\nonumber
\end{align}

\textbf{Neutrino Velocity Isocurvature:}
\begin{align}
\label{vcond}
&\delta_{c}|_{\tau=0}=\delta_{b}|_{\tau=0}=\delta_{\nu}|_{\tau=0}=\psi|_{\tau=0}=0\,,\\
&\text{Defining variable: }v_\nu.\nonumber
\end{align}

As with the adiabatic mode, similar conditions can be defined with other gauge invariant variables, such as the partial curvature perturbations $\zeta_s$. For example, a new set of conditions would be obtained simply by substituting every $\delta_s$ for the corresponding $\zeta_s$ and $\psi$ for the total $\zeta$. However, the new modes would not form a orthogonal basis in initial condition space, since choosing the $\zeta_s$ as defining variables would imply that the adiabatic mode contains a contribution from each of the density isocurvatures. The choice we present above is only one choice of variables which generate an orthogonal basis for the solution space. Many other choices are certainly possible, but this is the one used in the original literature \cite{Bucher:1999re}.

The conditions at second order are now already automatically set by stating that the Eqs.~\eqref{Adcond2}--\eqref{vcond} apply to the "non-perturbative" variables and not only to their first order parts. This is because, by definition, when we choose the component of the vector $X$ to be one of the defining variables, we have:
\be
\label{zero2nd}
I_i(\tau,k)=\sum_{j}{\mathcal{T}^i_j(\tau,k) I_j(k)}+\sum_{m,j}{\int_{k_1,k_2}\mathcal{T}^{i}_{mj}(\tau,k,k_1,k_2) I_m(k_1)I_j(k_2)}\,,
\ee
and thus, the obvious condition of equality, $I_i=I_i$, forces $\T^i_j=\delta^i_j$, as well as $\mathcal{T}^{i}_{mj}=0$, when the index $i$ corresponds to a defining variable. So, the condition is simply that the initial second order part of the defining variables is exactly zero, for all cases. The choice of defining variables plays a crucial role in the form of the results, as it determines which variables one chooses to be initially zero at second order. A different choice would result in equivalent results, but with a different functional form.

An additional condition must be set regarding the metric potential $E$. At linear order, the initial value of $E$ is not relevant for the evolution of the other quantities, but at second order, this is not the case, i.e. the first order $E|_{\tau=0}$ does appear in the quadratic source terms and would seem to influence the evolution. However, it can be shown that the initial condition of $E$ (or the value of $E$ at any one time point) can be fixed by the labelling of the spatial coordinates at that time point~\cite{Malik:2008im}. Therefore, it is fully consistent to set $E|_{\tau=0}=0$ and that is what we do throughout the paper.

With these conditions, one is now able to calculate the initial time evolution for the transfer functions in each mode. This will be done in the next section.

Before showing those results, a few important points must be made regarding the adiabatic nature of the second order modes. Firstly, it should be noted that, at second order, the different linear modes mix together. Thus, what we will later call the second order adiabatic mode is the one which is sourced by quadratic combinations of adiabatic linear modes only. Other modes exist which are sourced by one adiabatic component and another isocurvature one. We will label all those modes, \emph{mixed modes}. The second point is that, when this ``adiabatic mode" is defined in this way, it is not obvious that the entropy perturbation, given by
\begin{align}
S_{sr}^{(2)}=&\frac{\delta_s^{(2)}}{1+w_s}-\frac{\delta_r^{(2)}}{1+w_r}-\frac{2+w_s+w_r}{(1+w_s)^2}\left(\delta_s^{(1)}\right)^2\\\nonumber
&+\frac{2}{1+w_s}\delta_s^{(1)}\delta_r^{(1)}+\frac{2}{3(1+w_s)\mathcal{H}}\delta_s^{(1)}\left(\frac{\delta_s^{(1)\,\prime}}{1+w_s}-\frac{\delta_r^{(1)\,\prime}}{1+w_r}\right)\,,
\end{align}
should vanish at second order, since this condition was not enforced in any way. All the non-linear terms vanish since all the first order $\delta_i$ are initially zero when the mode is adiabatic. By the arguments following Eq.~\eqref{zero2nd}, we know that all second order densities are zero initially, except for the photon density, which is unconstrained by those arguments. However, the presence of a total density contrast can also be shown to generate decaying modes at second order. Therefore, since we are not considering decaying modes, by Eq.~\eqref{rhotot}, the photon density contrast is zero at second order as long as all first order velocities are zero. The mode considered here obeys this condition and is thus a true adiabatic mode.

In different gauges, the vanishing of the entropies may require different conditions for the density contrasts, particularly if they do not vanish initially at the linear level. For example, Ref.~\cite{Pettinari:2014vja} uses the following conditions, which should be valid in a general gauge, at second order:
\begin{gather}
\label{Adconds2}
\delta_{c}^{(2)}|_{\tau=0}=\delta_{b}^{(2)}|_{\tau=0}=\frac34\delta_{\gamma}^{(2)}|_{\tau=0}-\frac{3}{16}\left(\delta_{\gamma}^{(1)}\right)^2|_{\tau=0}\,,\delta_{\nu}^{(2)}|_{\tau=0}=\delta_{\gamma}^{(2)}|_{\tau=0}\,.
\end{gather}

Similar arguments apply to the isocurvature modes. Again, it is not obvious that the gauge invariant curvature perturbation, $\zeta$, will always vanish in all isocurvature modes, for the same reasons as above. For reference, in the large scale limit, $\zeta$ is given by
\begin{align}
\zeta^{(2)}=&-\psi^{(2)}+\frac{\delta^{(2)}}{3(1+w)}-\frac{1+3w}{9(1+w)^2}\left(\delta^{(1)}\right)^2\\\nonumber
&-\frac{4}{3(1+w)} \delta^{(1)}\psi^{(1)}+\frac{2}{3(1+w)\mathcal{H}}\delta^{(1)}\left(-\psi^{(1)\prime}+\frac{\delta^{(1)\prime}}{3 (1+w)}\right)\,,
\end{align}
where, for brevity, we are presenting only the variable which is invariant under changes of slicing. We can see that it depends only on the total density contrast, $\delta$, and not on the individual ones for each species. As explained above, $\delta$ is zero for growing modes, which added to the choice that $\psi|_{\tau=0}=0$ for isocurvature modes, results in $\zeta^{(2)}=0$, confirming also that all modes sourced only by isocurvatures are also true isocurvature modes.

\section{Approximate initial time evolution}\label{inievo}

In order to calculate the initial evolution for each mode, we expand every variable in powers of $\tau$:\footnote{To make this expansion well defined, one should use a dimensionless expansion parameter, instead of $\tau$, which has dimensions of time (or length, with $c=1$). In practice, as will be clear in the results, the expansion parameter will either be $k\tau$, $k_i\tau$ or $\omega\tau$, with $\omega\equiv\Omega_M \mathcal{H}/\sqrt{\Omega_R}$. The first two are very small for modes deep outside the horizon, while the last one is small for sufficiently early times, given that the constant $\omega$ is $O(10^{-6}) \text{Mpc}^{-1}$. Thus, the expansion in $\tau$ is correct as long as $\tau$ is sufficiently small.}
\be
X=X_0+X_1 \tau +X_2 \tau^2+X_3 \tau^3+...
\ee
This assumes we are neglecting decaying modes, as before. To find the solutions for each mode we apply one of the initial conditions given in Eqs.~\eqref{Adcond2}-\eqref{vcond} to the expansion of the variables $\{\psi,\delta_b,\delta_c,\delta_\nu,v_\nu\}$, generating a series of constraints on specific $X_I$. This constrained expansion is then substituted into the evolution equations, Eqs.~\eqref{EEq00}-\eqref{Boltz}, resulting in a set of algebraic equations for the coefficients, $X_I$, order by order in $\tau$. This will describe the initial solution to the equations of motion for each growing mode. We begin by applying this procedure at first order and recover the results found in Refs.~\cite{Bucher:1999re,Shaw:2009nf}. We substitute those results into the second order equations of motion and apply the same procedure to find the initial evolution for the second order transfer function. This is the final step to obtain our main results, which we show below.

We begin, however, by giving an example at linear order. We show here the results for the sum of the two matter isocurvature modes in synchronous gauge:
\begin{align}
\psi=&R_c\left(-\frac{1}{6}\omega\tau+\frac{1}{16}(\omega\tau)^2\right)\delta_c^0+R_b\left(-\frac{1}{6}\omega\tau+\frac{1}{16}(\omega\tau)^2\right)\delta_b^0\,,\nonumber\\
E=&\left(R_c\frac{15-4 R_\nu}{72(15+2 R_\nu)}\omega\tau^3\right)\delta_c^0+\left(R_b\frac{15-4 R_\nu}{72(15+2 R_\nu)}\omega\tau^3\right)\delta_b^0\,,\nonumber\\
\delta_{c}=&\left(1-\frac{R_c}{2}\omega\tau+\frac{3R_c}{16}(\omega\tau)^2\right)\delta_c^0+R_b\left(-\frac{1}{2}\omega\tau+\frac{3}{16}(\omega\tau)^2\right)\delta_b^0\,,\nonumber\\
\delta_{b}=&\left(-\frac{R_c}{2}\omega\tau+\frac{3R_c}{16}(\omega\tau)^2\right)\delta_c^0+\left(1-\frac{R_b}{2}\omega\tau+\frac{3R_b}{16}(\omega\tau)^2\right)\delta_b^0\,,\nonumber\\
\delta_{\gamma}=&\left(-\frac{2R_c}{3}\omega\tau+\frac{R_c}{4}(\omega\tau)^2\right)\delta_c^0+\left(-\frac{2R_b}{3}\omega\tau+\frac{R_b}{4}(\omega\tau)^2\right)\delta_b^0\,,\nonumber\\
\delta_{\nu}=&\left(-\frac{2R_c}{3}\omega\tau+\frac{R_c}{4}(\omega\tau)^2\right)\delta_c^0+\left(-\frac{2R_b}{3}\omega\tau+\frac{R_b}{4}(\omega\tau)^2\right)\delta_b^0\,,\nonumber\\
v_{\gamma b}=&\left(\frac{R_c }{12}\omega\tau^2\right)\delta_c^0+\left(\frac{R_b }{12}\omega\tau^2\right)\delta_b^0\,,\nonumber\\
v_{\nu}=&\left(\frac{R_c }{12}\omega\tau^2\right)\delta_c^0+\left(\frac{R_b }{12}\omega\tau^2\right)\delta_b^0\,,\label{matteriso1}\\
\sigma_\nu=&\left(-\frac{R_c}{6(15+2 R_\nu)}k^2\omega\tau^3\right)\delta_c^0+\left(-\frac{R_b}{6(15+2 R_\nu)}k^2\omega\tau^3\right)\delta_b^0\,,\nonumber
\end{align}
in which $\omega\equiv\Omega_M \mathcal{H}/\sqrt{\Omega_R}$, $R_c=\Omega_c/\Omega_M$, $R_\nu=\Omega_\nu/\Omega_R$, $R_\gamma=\Omega_\gamma/\Omega_R$ and the $\Omega_s$ are the usual density parameters. We have also used the total matter and total radiation density parameters, respectively given by $\Omega_M=\Omega_c+\Omega_b$ and $\Omega_R=\Omega_\gamma+\Omega_\nu$. This implies that $R_c+R_b=1$ as well as $R_\nu+R_\gamma=1$. We have also abbreviated the initial values of the cold dark matter and baryon density contrasts, $\delta_c|_{\tau=0}$ and $\delta_b|_{\tau=0}$, to $\delta_c^0$ and $\delta_b^0$ for simplicity of notation. We also do this for all other defining variables in all modes presented below.

This example is particularly useful because it also allows us to analyse a combination of modes called the compensated isocurvature mode~\cite{Grin:2011tf}. This mode is defined by the choice of initial conditions for which all variables cancel in the equations above, except the matter density contrasts. It is given by the following condition
\begin{equation}
\label{CIP}
\delta_b^0=-\frac{R_c}{R_b}\delta_c^0\,.
\end{equation}
When the initial conditions are exactly related in this way, no other variables are generated at linear order. As we will later verify, this is no longer true at second order, due to mode mixing.

Another property that we can see in this example is that, at first order in perturbation theory, there is a hierarchy between the brightness tensors in terms of their order in $\tau$: it is clear here, that $\delta_\nu\gg v_\nu\gg \sigma_\nu$. This can be shown using the evolution equations for those variables --- the first order versions of Eqs.~\eqref{Eqvnu2}--\eqref{Boltz} --- from which one deduces that $v_\nu\propto \int \delta_\nu \rm{d}\tau$ and $\sigma_\nu\propto \int v_\nu\rm{d}\tau$. This implies that one can safely neglect the higher rank brightness tensors, as they will certainly be smaller than the ones shown. At second order, this is not so straightforward, as all variables are sourced by non-linear terms, which do not have to obey such a hierarchy. In order to test this, all the results below include one extra variable, the scalar part of the rank 3 brightness tensor, $\Delta_3$. Should this variable be of the same order in $\tau$ as $\sigma_\nu$, one may assume that all other brightness tensors are of a similar size. Should that be the case, they may not be negligible, since they may affect the evolution of all other variables. In practice, as we show below, none of the modes under study suffer from this problem and this hierarchy is preserved.

We now present the second order results for all growing modes, excluding the neutrino velocity mode, as it includes decaying contributions at second order, as discussed above. In all of the results shown, we abuse the notation and use the names of the variables to denote the transfer functions multiplied by the defining variables (for example $\psi^{(2)}=\T_{ij}I_iI_j$) i.e. we show only the integrand of the second order part of the variable. We begin by showing the pure adiabatic mode and show the results for the isocurvature modes after that by ``activating" each of the four linear growing modes separately.

\subsection{Pure adiabatic mode}

We find the following results for the initial evolution at second order and at leading order in $\tau$, when including only the quadratic source composed by the adiabatic first order solutions, in synchronous gauge:
\begin{align}
\psi^{(2)}=&-\frac{4R_\nu k^2(3k^2+k_1^2+k_2^2)+5\left(3(k_1^2-k_2^2)^2+k^2(k_1^2+k_2^2)\right)}{24(4R_\nu+15) k^4}(k\tau)^2\psi^{0}_{k_1}\psi^{0}_{k_2}\,,\nonumber\\
E^{(2)}=&-\frac{5\left(9k^4-3(k_1^2-k_2^2)^2+2k^2(k_1^2+k_2^2)\right)}{8(4R_\nu+15) k^4}\tau^2\psi^{0}_{k_1}\psi^{0}_{k_2}\,,\nonumber\\
\delta_{c}^{(2)}=&-\frac{1}{8}\left(3 k^2+5(k_1^2+k_2^2)\right)\tau^2\psi^{0}_{k_1}\psi^{0}_{k_2}\,,\nonumber\\
\delta_{b}^{(2)}=&-\frac{1}{8}\left(3 k^2+5(k_1^2+k_2^2)\right)\tau^2\psi^{0}_{k_1}\psi^{0}_{k_2}\,,\nonumber\\
\delta_{\gamma}^{(2)}=&-\frac{1}{6}\left(3 k^2+5(k_1^2+k_2^2)\right)\tau^2\psi^{0}_{k_1}\psi^{0}_{k_2}\,,\nonumber\\
\delta_{\nu}^{(2)}=&-\frac{1}{6}\left(3 k^2+5(k_1^2+k_2^2)\right)\tau^2\psi^{0}_{k_1}\psi^{0}_{k_2}\,,\\
v_{\gamma b}^{(2)}=&\frac{1}{72k^2}\left(3k^4+2(k_1^2-k_2^2)^2+7k^2(k_1^2+k_2^2)\right)\tau^3\psi^{0}_{k_1}\psi^{0}_{k_2}\,,\nonumber\\
v_{\nu}^{(2)}=&\frac{23+4R_\nu}{72(4R_\nu+15)k^2}\left(3k^4+2(k_1^2-k_2^2)^2+7k^2(k_1^2+k_2^2)\right)\tau^3\psi^{0}_{k_1}\psi^{0}_{k_2}\,,\nonumber\\
\sigma_\nu^{(2)}=&\frac{\left(9k^4-3(k_1^2-k_2^2)^2+2k^2(k_1^2+k_2^2)\right)}{6(4R_\nu+15) k^4}(k\tau)^2\psi^{0}_{k_1}\psi^{0}_{k_2}\,,\nonumber\\
\Delta_3^{(2)}=&-\frac{37k^4+9(k_1^2-k_2^2)^2-6k^2(k_1^2+k_2^2)}{42(15+4R_\nu)k^4}\tau^3\psi^{0}_{k_1}\psi^{0}_{k_2}\,,\nonumber
\end{align}

These results for the adiabatic mode were already known in Poisson gauge \cite{Pettinari:2014vja,Pitrou:2010sn} and one can check that they match ours by using the gauge transformations given in Appendix \ref{gaugetr}. We see here that $\sigma_\nu$ is initially larger (in order of $\tau$) than $v_\nu$. This was not the case at the linear level. However, we also note that $\Delta_3$ is again higher order in $\tau$, giving us confidence that higher rank tensors can be neglected.

\subsection{Pure cold dark matter isocurvature mode}

For the mode that is sourced by the quadratic dark matter isocurvature first order solutions, the initial evolution is given by:
\begin{align}
\psi^{(2)}=&R_c^2\left(\frac{(\omega\tau)^2}{48}-\frac{(\omega\tau)^3}{72}\right)\delta_{c,k_1}^{0}\delta_{c,k_2}^{0}\,,\nonumber\\
E^{(2)}=&O(\tau^4)\nonumber\\
\delta_{c}^{(2)}=&R_c\left(-\omega\tau+\frac{18+23R_c}{48}(\omega\tau)^2+\frac{16 (k_1^2+k_2^2)-15(6+17R_c)\omega^2}{720}\omega\tau^3\right)\delta_{c,k_1}^{0}\delta_{c,k_2}^{0}\,,\nonumber\\
\delta_{b}^{(2)}=&R_c^2\left(\frac{23}{48}(\omega\tau)^2-\frac{17}{48}(\omega\tau)^3\right)\delta_{c,k_1}^{0}\delta_{c,k_2}^{0}\,,\nonumber\\
\delta_{\gamma}^{(2)}=&R_c^2\left(\frac{3}{4}(\omega\tau)^2-\frac59(\omega\tau)^3\right)\delta_{c,k_1}^{0}\delta_{c,k_2}^{0}\,,\nonumber\\
\delta_{\nu}^{(2)}=&R_c^2\left(\frac{3}{4}(\omega\tau)^2-\frac59(\omega\tau)^3\right)\delta_{c,k_1}^{0}\delta_{c,k_2}^{0}\,,\\
v_{\gamma b}^{(2)}=&R_c^2\left(-\frac{7\omega^2\tau^3}{144}+\frac{(15 R_b+16 R_\gamma)\omega^3\tau^4}{576R_\gamma}\right)\delta_{c,k_1}^{0}\delta_{c,k_2}^{0}\,,\nonumber\\
v_{\nu}^{(2)}=&R_c^2\left(-\frac{7\omega^2\tau^3}{144}+\frac{\omega^3\tau^4}{36}\right)\delta_{c,k_1}^{0}\delta_{c,k_2}^{0}\,,\nonumber\\
\sigma_\nu^{(2)}=&O(\tau^4)\,,\nonumber\\
\Delta_3^{(2)}=&O(\tau^5)\,.\nonumber
\end{align}

\subsection{Mixture of adiabatic and cold dark matter modes}

When both the adiabatic mode and the dark matter isocurvature are present, a mixed mode is generated, for which the initial evolution is:
\begin{align}
\psi^{(2)}=&R_c\left(\frac{1}{3}\omega\tau-\frac{1}{8}(\omega\tau)^2\right)\delta_{c,k_1}^{0}\psi^{0}_{k_2}\,,\nonumber\\
E^{(2)}=&f_E^{c\psi}(k,k_1,k_2)\omega\tau^3\delta_{c,k_1}^{0}\psi^{0}_{k_2}\nonumber\\
\delta_{c}^{(2)}=&\left(-\frac{1}{4}k_2^2\tau^2 +\frac{1}{180} (-2(k^2-5k_1^2)R_c +k_2^2(9+41R_c))\omega\tau^3\right)\delta_{c,k_1}^{0}\psi^{0}_{k_2}\,,\nonumber\\
\delta_{b}^{(2)}=&-\frac{R_c}{120}\omega\tau^3(3k^2-15k_1^2-29k_2^2)\delta_{c,k_1}^{0}\psi^{0}_{k_2}\,,\nonumber\\
\delta_{\gamma}^{(2)}=&-\frac{R_c}{90}\omega\tau^3(3k^2-15k_1^2-34k_2^2)\delta_{c,k_1}^{0}\psi^{0}_{k_2}\,,\nonumber\\
\delta_{\nu}^{(2)}=&-\frac{R_c}{90}\omega\tau^3(3k^2-15k_1^2-34k_2^2)\delta_{c,k_1}^{0}\psi^{0}_{k_2}\,,\\
v_{\gamma b}^{(2)}=&\left(\frac{R_c}{12k^2}(k^2+k_1^2-k_2^2)\omega\tau^2-\frac{R_c(R_\gamma+3R_b)}{48R_\gamma k^2}(k^2+k_1^2-k_2^2)\omega^2\tau^3\right)\delta_{c,k_1}^{0}\psi^{0}_{k_2}\,,\nonumber\\
v_{\nu}^{(2)}=&\left(\frac{R_c}{12k^2}(k^2+k_1^2-k_2^2)\omega\tau^2-\frac{R_c}{48k^2}(k^2+k_1^2-k_2^2)\omega^2\tau^3\right)\delta_{c,k_1}^{0}\psi^{0}_{k_2}\,,\nonumber\\
\sigma_\nu^{(2)}=&f^{c\psi}_\sigma(k,k_1,k_2)\omega k^2\tau^3\delta_{c,k_1}^{0}\psi^{0}_{k_2}\,,\nonumber\\
\Delta_3^{(2)}=&O(\tau^4)\,,\nonumber
\end{align}
with the following kernels:
\begin{align}
f^{c\psi}_E=&-\frac{R_c}{576 (15+4R_\nu) (15+2R_\nu) k^4}\left[(225 + 720 R_\nu + 32 R_\nu^2) k^4\nonumber\right.\\
&+ 3 (675 + 240 R_\nu - 32 R_\nu^2) (k_1^2 - k_2^2)^2 \nonumber\\
&\left. +  2 k^2 ((-1125 - 720 R_\nu + 32 R_\nu^2) k_1^2 + (-225 + 240 R_\nu + 32 R_\nu^2) k_2^2)\right]\,,\nonumber\\
f^{c\psi}_\sigma=&-\frac{R_c\left[(135 + 8 R_\nu) k^4 +3 (5 - 8 R_\nu) (k_1^2 - k_2^2)^2 + \nonumber
 2 k^2 ((-75 + 8 R_\nu) k_1^2 + (65 + 8 R_\nu) k_2^2)\right]}{48 (15+4R_\nu) (15+2R_\nu) k^4}\,.
\end{align}
Note that to get the full results for the mixed mode one would have to add the complementary solution obtained by switching $k_1\leftrightarrow k_2$. We can see that there exist growing mode solutions for these mixed modes, thus showing that they must be taken into account if one is to have an accurate understanding of the effect of isocurvature modes on non-linear observables. This is even more important in the particular case shown, since this mode includes a contribution from the adiabatic mode, which should make this mixed mode more relevant than the pure isocurvature one, presented before.

\subsection{Pure baryon isocurvature mode}

We now move on to the introduction of the baryon isocurvature mode:
\begin{align}
\psi^{(2)}=&R_b^2\left(\frac{(\omega\tau)^2}{48}-\frac{(\omega\tau)^3}{72}\right)\delta_{b,k_1}^{0}\delta_{b,k_2}^{0}\,,\nonumber\\
E^{(2)}=&O(\tau^4)\nonumber\\
\delta_{c}^{(2)}=&R_b^2\left(\frac{23}{48}(\omega\tau)^2-\frac{17}{48}(\omega\tau)^3\right)\delta_{b,k_1}^{0}\delta_{b,k_2}^{0}\,,\nonumber\\
\delta_{b}^{(2)}=&R_b\omega\tau\left(-1+\frac{18+23R_b}{48}\omega\tau+\frac{16 (k_1^2+k_2^2)+20k^2-15(6+17R_c)\omega^2}{720}\tau^2\right)\delta_{b,k_1}^{0}\delta_{b,k_2}^{0}\,,\nonumber\\
\delta_{\gamma}^{(2)}=&R_b^2\left(\frac{3}{4}(\omega\tau)^2-\frac59(\omega\tau)^3\right)\delta_{b,k_1}^{0}\delta_{b,k_2}^{0}\,,\nonumber\\
\delta_{\nu}^{(2)}=&R_b^2\left(\frac{3}{4}(\omega\tau)^2-\frac59(\omega\tau)^3\right)\delta_{b,k_1}^{0}\delta_{b,k_2}^{0}\,,\\
v_{\gamma b}^{(2)}=&R_b^2\left(\frac{7R_\nu-16}{144R_\gamma}\omega^2\tau^3+\frac{R_b(69-15R_\nu)+16R_\gamma^2}{576R_\gamma^2}\omega^3\tau^4\right)\delta_{b,k_1}^{0}\delta_{b,k_2}^{0}\,,\nonumber\\
v_{\nu}^{(2)}=&R_b^2\left(-\frac{7}{144}\omega^2\tau^3+\frac{1}{36}\omega^3\tau^4\right)\delta_{b,k_1}^{0}\delta_{b,k_2}^{0}\,,\nonumber\\
\sigma_\nu^{(2)}=&O(\tau^4)\,,\nonumber\\
\Delta_3^{(2)}=&O(\tau^5)\,.\nonumber
\end{align}
This mode is very similar to the ``pure" dark matter isocurvature, as it is already at first order. In this case, however, the application of the compensated isocurvature condition, Eq.~\eqref{CIP}, would not lead to cancellations when this mode is summed to the dark matter one, due the quadratic nature of these modes. Furthermore, some terms are completely different in the two modes, namely the matter densities and the baryon-photon velocity. However, in order to completely analyse the initial evolution of the compensated isocurvature mode, we must still investigate the mixed mode between the baryon and dark matter isocurvatures. 

\subsection{Mixture of baryon and cold dark matter modes}

This mixed mode is given by
\begin{align}
\psi^{(2)}=&\frac{R_bR_c}{48}\left((\omega\tau)^2-\frac23 (\omega\tau)^3\right)\delta_{b,k_1}^{0}\delta_{c,k_2}^{0}\,,\nonumber\\
E^{(2)}=&O(\tau^4)\nonumber\\
\delta_{c}^{(2)}=&R_b\left(-\frac{1}{2}\omega\tau+\frac{9+23 R_c}{48}(\omega\tau)^2 + \frac{16 k_1^2 - 15 (3 + 17 R_c) \omega^2}{720} \omega \tau^3\right)\delta_{b,k_1}^{0}\delta^{0}_{c,k_2}\,,\nonumber\\
\delta_{b}^{(2)}=&R_c\left(-\frac{1}{2}\omega\tau+\frac{9+23 R_b}{48}(\omega\tau)^2 + \frac{10 k^2 - 10 k_1^2 + 26 k_2^2 - 300 \omega^2 + 255 R_c \omega^2}{720} \omega \tau^3 \right)\delta_{b,k_1}^{0}\delta^{0}_{c,k_2}\,,\nonumber\\
\delta_{\gamma}^{(2)}=&R_bR_c\left(\frac{3}{4}(\omega\tau)^2-\frac{5}{9}(\omega\tau)^3\right)\delta_{b,k_1}^{0}\delta_{c,k_2}^{0}\,,\nonumber\\
\delta_{\nu}^{(2)}=&R_bR_c\left(\frac{3}{4}(\omega\tau)^2-\frac{5}{9}(\omega\tau)^3\right)\delta_{b,k_1}^{0}\delta_{c,k_2}^{0}\,,\\
v_{\gamma b}^{(2)}=&\frac{R_bR_c(9(k_1^2-k_2^2)-(23-14R_\nu)k^2)}{288R_\gamma k^2}\omega^2\tau^3\delta_{b,k_1}^{0}\delta_{c,k_2}^{0}\,,\nonumber\\
v_{\nu}^{(2)}=&-\frac{7R_bR_c}{144}\omega^2\tau^3\delta_{b,k_1}^{0}\delta_{c,k_2}^{0}\,,\nonumber\\
\sigma_\nu^{(2)}=&O(\tau^4)\,,\nonumber\\
\Delta_3^{(2)}=&O(\tau^5)\,.\nonumber
\end{align}
Adding all the matter modes together and applying the compensated isocurvature condition, Eq.~\eqref{CIP}, we can show that again, the compensated isocurvature mode has vanishing initial evolution even at second order. This is not surprising, since, if only these matter isocurvature modes are active and do not evolve at linear order, they would only source the second order evolution if terms like $\delta_c^2$, $\delta_b^2$ or $\delta_c\delta_b$ existed in the evolution equations. Having concluded that a pure compensated isocurvature mode does not evolve initially, it remains to be seen whether it can mix with the adiabatic mode and generate additional contributions. 

\subsection{Mixture of adiabatic and baryon modes}

To test what happens when one mixes a compensated isocurvature with the adiabatic mode, we first need the mixed mode between the baryon isocurvature and the adiabatic mode:
\begin{align}
\psi^{(2)}=&R_b\left(\frac{1}{3}\omega\tau-\frac{1}{8}\omega^2\tau^2\right)\delta_{b,k_1}^{0}\psi^{0}_{k_2}\,,\nonumber\\
E^{(2)}=&f_E^{b\psi}(k,k_1,k_2)\omega\tau^3\delta_{b,k_1}^{0}\psi^{0}_{k_2}\,,\nonumber\\
\delta_{c}^{(2)}=&-\frac{R_b}{180}\omega\tau^3(-2k^2+10k_1^2+41k_2^2)\delta_{b,k_1}^{0}\psi^{0}_{k_2}\,,\nonumber\\
\delta_{b}^{(2)}=&\left(-\frac{1}{4}k_2^2\tau^2 +\frac{1}{120} ((15k_1^2+29k_2^2-3k^2)R_b +6 k_2^2)\omega\tau^3\right)\delta_{b,k_1}^{0}\psi^{0}_{k_2}\,,\nonumber\\
\delta_{\gamma}^{(2)}=&-\frac{R_b}{90}\omega\tau^3(3k^2-15k_1^2-34k_2^2)\delta_{b,k_1}^{0}\psi^{0}_{k_2}\,,\nonumber\\
\delta_{\nu}^{(2)}=&-\frac{R_b}{90}\omega\tau^3(3k^2-15k_1^2-34k_2^2)\delta_{b,k_1}^{0}\psi^{0}_{k_2}\,,\\
v_{\gamma b}^{(2)}=&\left(\frac{R_b}{12k^2}(k^2+k_1^2-k_2^2)\omega\tau^2-\frac{R_b(R_\gamma+3R_b)}{48R_\gamma k^2}(k^2+k_1^2-k_2^2)\omega^2\tau^3\right)\delta_{b,k_1}^{0}\psi^{0}_{k_2}\,,\nonumber\\
v_{\nu}^{(2)}=&\left(\frac{R_b}{12k^2}(k^2+k_1^2-k_2^2)\omega\tau^2-\frac{R_b}{48k^2}(k^2+k_1^2-k_2^2)\omega^2\tau^3\right)\delta_{b,k_1}^{0}\psi^{0}_{k_2}\,,\nonumber\\
\sigma_\nu^{(2)}=&f^{b\psi}_\sigma(k,k_1,k_2)\omega k^2\tau^3\delta_{b,k_1}^{0}\psi^{0}_{k_2}\,,\nonumber\\
\Delta_3^{(2)}=&O(\tau^4)\,,\nonumber
\end{align}
with the following kernels:
\begin{gather}
f^{b\psi}_E(k,k_1,k_2)=\frac{R_b}{R_c}f^{c\psi}_E(k,k_1,k_2)\,,\nonumber
\\
f^{b\psi}_\sigma(k,k_1,k_2)=\frac{R_b}{R_c}f^{c\psi}_\sigma(k,k_1,k_2)\,.\nonumber
\end{gather}
It is immediately clear from the relationship between the kernels for $E$ and $\sigma$, that cancellations will occur when the compensated isocurvature condition, Eq.~\eqref{CIP}, is applied. However, there are some terms that do survive and are given by
\begin{align}
\delta_{c}^{(2)}=&-\frac{1}{20} k_2^2\tau^2(5-\omega\tau)\delta_{\text{CI},k_1}^{0}\psi^{0}_{k_2}\,,\nonumber\\
\delta_{b}^{(2)}=&\frac{R_c}{20R_b} k_2^2\tau^2(5-\omega\tau)\delta_{\text{CI},k_1}^{0}\psi^{0}_{k_2}\,,\\
v_{\gamma b}^{(2)}=&\frac{R_c}{R_\gamma}\frac{k^2+k_1^2-k_2^2}{96k^2}k_2^2\omega\tau^4\delta_{\text{CI},k_1}^{0}\psi^{0}_{k_2}\,,\nonumber
\end{align}
in which $\delta_{\text{CI},k_1}^{0}$ is the initial density contrast of dark matter in the compensated isocurvature mode. We see here that the compensated isocurvature condition is conserved, i.e. $\delta_{b}^{(2)}=-\frac{R_c}{R_b}\delta_c^{(2)}$, but we also see that the velocity fluctuation of the baryon-photon plasma is generated by this mode, which was non-existent at linear order. We confirm here that the compensated isocurvature mode does have an effect on the evolution at second order, even at these early times.

\subsection{Pure neutrino density isocurvature mode}

We now introduce the modes sourced by the neutrino density isocurvature. First we show the results for the ``pure" mode:
\begin{align}
\psi^{(2)}=&f^{\nu\nu}_\psi(k,k_1,k_2) (k\tau)^2\delta_{\nu,k_1}^{0}\delta_{\nu,k_2}^{0}\,,\nonumber\\
E^{(2)}=&f^{\nu\nu}_E(k,k_1,k_2) \tau^2\delta_{\nu,k_1}^{0}\delta_{\nu,k_2}^{0}\,,\nonumber\\
\delta_{c}^{(2)}=&-\frac{R_b R_\nu^2}{320R_\gamma^2}(7k^2-3(k_1^2+k_2^2))\omega\tau^3\delta_{\nu,k_1}^{0}\delta_{\nu,k_2}^{0}\,,\nonumber\\
\delta_{b}^{(2)}=&\frac{R_\nu^2}{32R_\gamma^2}(7k^2-3(k_1^2+k_2^2))\tau^2\delta_{\nu,k_1}^{0}\delta_{\nu,k_2}^{0}\,,\nonumber\\
\delta_{\gamma}^{(2)}=&\frac{R_\nu^2}{12R_\gamma^2}(k^2-k_1^2-k_2^2)\tau^2\delta_{\nu,k_1}^{0}\delta_{\nu,k_2}^{0}\,,\nonumber\\
\delta_{\nu}^{(2)}=&\frac{1}{12}(k^2-k_1^2-k_2^2)\tau^2\delta_{\nu,k_1}^{0}\delta_{\nu,k_2}^{0}\,,\\
v_{\gamma b}^{(2)}=&\left(\frac{R_\nu^2}{4R_\gamma^2}\tau-\frac{3R_b R_\nu^2}{8R_\gamma^3}\omega\tau^2\right)\delta_{\nu,k_1}^{0}\delta_{\nu,k_2}^{0}\,,\nonumber\\
v_{\nu}^{(2)}=&\frac{1}{4}\tau\delta_{\nu,k_1}^{0}\delta_{\nu,k_2}^{0}\,,\nonumber\\
\sigma_\nu^{(2)}=&f^{\nu\nu}_\sigma(k,k_1,k_2)(k\tau)^2\delta_{\nu,k_1}^{0}\delta_{\nu,k_2}^{0}\,,\nonumber\\
\Delta_3^{(2)}=&f^{\nu\nu}_\Delta(k,k_1,k_2)\tau^3\delta_{\nu,k_1}^{0}\delta^{0}_{\nu,k_2}\,,\nonumber
\end{align}
in which the kernels abbreviated above are given by
\begin{align}
f^{\nu\nu}_\psi(k,k_1,k_2)=&-\frac{R_\nu^2\left[(27 + 68 R_\nu) k^4 -(91 + 4 R_\nu) \left( 3 (k_1^2 - k_2^2)^2 - 2 k^2 (k_1^2 + k_2^2)\right)\right]}{96R_\gamma(4R_\nu+15)^2 k^4}\,,\nonumber\\
f^{\nu\nu}_E(k,k_1,k_2)=&-3 f^{\nu\nu}_\psi(k,k_1,k_2)\,,\nonumber\\
f^{\nu\nu}_\sigma(k,k_1,k_2)=&-\frac{1}{96R_\gamma(4R_\nu+15)^2 k^4}\left[(-225 - 39 R_\nu + 188 R_\nu^2) k^4 \right.\nonumber\\
&\left.+ (225 - 153 R_\nu + 4 R_\nu^2)\left(3  (k_1^2 - k_2^2)^2 - 2  k^2 (k_1^2 + k_2^2)\right)\right]\nonumber\,,\\
f^{\nu\nu}_\Delta(k,k_1,k_2)=&-\frac{R_\nu\left[(-51 + 32 R_\nu) k^4 + (3 + 16 R_\nu)\left(3  (k_1^2 - k_2^2)^2 - \nonumber
 2  k^2 (k_1^2 + k_2^2)\right)\right]}{84R_\gamma(4R_\nu+15)^2 k^4}\,.
\end{align}

\subsection{Mixture of adiabatic and neutrino modes}

The mixed mode between the neutrino density isocurvature and the adiabatic mode is given by
\begin{align}
\psi^{(2)}=&f_\psi^{\nu\psi}(k,k_1,k_2)(k\tau)^2\delta_{\nu,k_1}^{0}\psi^{0}_{k_2}\,,\nonumber\\
E^{(2)}=&f_E^{\nu\psi}(k,k_1,k_2)\tau^2\delta_{\nu,k_1}^{0}\psi^{0}_{k_2}\,,\nonumber\\
\delta_{c}^{(2)}=&\frac{R_bR_\nu}{160R_\gamma}(k^2-5k_1^2-k_2^2)\omega\tau^3\delta_{b,k_1}^{0}\psi^{0}_{k_2}\,,\nonumber\\
\delta_{b}^{(2)}=&-\frac{R_\nu}{16R_\gamma}(k^2-5k_1^2-k_2^2)\tau^2\delta_{\nu,k_1}^{0}\psi^{0}_{k_2}\,,\nonumber\\
\delta_{\gamma}^{(2)}=&-\frac{R_\nu}{12R_\gamma}(k^2-5(k_1^2+k_2^2))\tau^2\delta_{\nu,k_1}^{0}\psi^{0}_{k_2}\,,\nonumber\\
\delta_{\nu}^{(2)}=&\frac{1}{12}(k^2-5(k_1^2+k_2^2))\tau^2\delta_{\nu,k_1}^{0}\psi^{0}_{k_2}\,,\\
v_{\gamma b}^{(2)}=&\left(\frac{R_\nu(k^2+k_1^2-k_2^2)}{4R_\gamma k^2}\tau-\frac{3R_bR_\nu(k^2+k_1^2-k_2^2)}{16R_\gamma^2k^2}\omega\tau^2\right)\delta_{\nu,k_1}^{0}\psi^{0}_{k_2}\,,\nonumber\\
v_{\nu}^{(2)}=&-\frac{(k^2+k_1^2-k_2^2)}{4k^2}\tau\delta_{\nu,k_1}^{0}\psi^{0}_{k_2}\,,\nonumber\\
\sigma_\nu^{(2)}=&f_\sigma^{\nu\psi}(k,k_1,k_2) (k\tau)^2\delta_{\nu,k_1}^{0}\psi^{0}_{k_2}\,,\nonumber\\
\Delta_3^{(2)}=&f^{\nu\psi}_\Delta(k,k_1,k_2)\tau^3\delta_{\nu,k_1}^{0}\psi^{0}_{k_2}\,.\nonumber
\end{align}
The kernels are given by
\begin{align}
f^{\nu\psi}_\psi=&-\frac{R_\nu\left[(45 + 4 R_\nu) k^4 - 3 (5 + 4 R_\nu) (k_1^2 - k_2^2)^2 + 
 k^2 ((-30 + 8 R_\nu) k_1^2 + 2 (25 + 4 R_\nu) k_2^2)\right]}{24(4R_\nu+15)^2 k^4}\,,\nonumber\\
f^{\nu\psi}_E=&-3f^{\nu\psi}_\psi\,,\nonumber\\
f^{\nu\psi}_\sigma=&-\frac{3}{R_\nu}f^{\nu\psi}_\psi(k,k_1,k_2)\,,\nonumber\\
f^{\nu\psi}_\Delta=&-\frac{1}{336(15+4R\nu)^2 k^6}\left[(1545 + 316 R_\nu) k^6 + 35 (15 + 4 R_\nu) (k_1^2 - k_2^2)^3\right.\nonumber\\
&\left. - 3 k^2 (k_1^2 - k_2^2) (3 (65 + 28 R_\nu) k_1^2 + (225 + 28 R_\nu) k_2^2) + k^4 ((675 + 372 R_\nu) k_1^2 - 5 (147 + 52 R_\nu) k_2^2)\right]\nonumber\,.
\end{align}

\subsection{Mixture of dark matter and neutrino modes}

Now we show the neutrino-dark matter mixed mode:
\begin{align}
\psi^{(2)}=&f_\psi^{\nu c}(k,k_1,k_2)\omega k^2\tau^3\delta_{\nu,k_1}^{0}\delta_{c,k_2}^{0}\,,\nonumber\\
E^{(2)}=&f_E^{\nu c}(k,k_1,k_2)\omega\tau^3\delta_{\nu,k_1}^{0}\delta_{c,k_2}^{0}\,,\nonumber\\
\delta_{c}^{(2)}=&-\frac{R_bR_\nu}{80R_\gamma}k_1^2\omega\tau^3\delta_{\nu,k_1}^{0}\delta_{c,k_2}^{0}\,,\nonumber\\
\delta_{b}^{(2)}=&-\frac{R_\nu R_c}{288R_\gamma}(-5k^2+29k_1^2+5k_2^2)\omega\tau^3\delta_{\nu,k_1}^{0}\delta_{c,k_2}^{0}\,,\nonumber\\
\delta_{\gamma}^{(2)}=&\frac{R_\nu R_c}{R_\gamma}\left(\frac{2}{3}\omega\tau-\frac{1}{4}(\omega\tau)^2\right)\delta_{\nu,k_1}^{0}\delta_{c,k_2}^{0}\,,\nonumber\\
\delta_{\nu}^{(2)}=&\left(-\frac{2R_c}{3}\omega\tau+\frac{R_c}{4}(\omega\tau)^2\right)\delta_{\nu,k_1}^{0}\delta_{c,k_2}^{0}\,,\\
v_{\gamma b}^{(2)}=&-\frac{R_\nu R_c}{R_\gamma}\left(\frac{k^2+k_1^2-k_2^2}{32k^2}\omega\tau^2+\frac{k^2(9R_b-4R_\gamma)-(k_1^2-k_2^2)(4R_\gamma+15 R_b)}{384R_\gamma k^2}\omega^2\tau^3\right)\delta_{\nu,k_1}^{0}\delta_{c,k_2}^{0}\,,\nonumber\\
v_{\nu}^{(2)}=&\left(\frac{(k^2+k_1^2-k_2^2)R_c}{32k^2}\omega\tau^2-\frac{(k^2+k_1^2-k_2^2)R_c}{96k^2}\omega^2\tau^3\right)\delta_{\nu,k_1}^{0}\delta_{c,k_2}^{0}\,,\nonumber\\
\sigma_\nu^{(2)}=&f_\sigma^{\nu c}(k,k_1,k_2) \omega k^2 \tau^3\delta_{\nu,k_1}^{0}\delta_{c,k_2}^{0}\,,\nonumber\\
\Delta_3^{(2)}=&O(\tau^4)\,,\nonumber
\end{align}
with the following kernels:
\begin{align}
f^{\nu c}_\psi(k,k_1,k_2)=&\frac{R_\nu R_c}{144(2R_\nu+15)^2(4R_\nu+15) k^4}\left[(675 + 90 R_\nu - 6 R_\nu^2) (k^4+ (k_1^2 - k_2^2)^2)\right.\nonumber\\
&\left. - 2 k^2 ((225 + 90 R_\nu + 2 R_\nu^2) k_1^2 - 3 (-75 + 10 R_\nu + 2 R_\nu^2) k_2^2)\right]\nonumber\,,\\
f^{\nu c}_E(k,k_1,k_2)=&-3f^{\nu c}_\psi(k,k_1,k_2)\,,\nonumber\\
f^{\nu c}_\sigma(k,k_1,k_2)=&\frac{R_c}{96(2R_\nu+15)^2(4R_\nu+15) k^4}\left[3 (-1125 - 180 R_\nu + 4 R_\nu^2) (k^4+(k_1^2 - k_2^2)^2) \right.\nonumber\\
&\left. + 2 k^2 ((675 + 300 R_\nu + 4 R_\nu^2) k_1^2 +  3 (525 + 20 R_\nu - 4 R_\nu^2) k_2^2)\right]\nonumber\,.
\end{align}

\subsection{Mixture of baryon and neutrino modes}

Finally, the results for the neutrino-baryon mixed mode are
\begin{align}
\psi^{(2)}=&f_\psi^{\nu b}(k,k_1,k_2)\omega k^2\tau^3\delta_{\nu,k_1}^{0}\delta_{b,k_2}^{0}\,,\nonumber\\
E^{(2)}=&f_E^{\nu b}(k,k_1,k_2)\omega\tau^3\delta_{\nu,k_1}^{0}\delta_{b,k_2}^{0}\,,\nonumber\\
\delta_{c}^{(2)}=&-\frac{R_bR_\nu}{160R_\gamma}(k^2+k_1^2-k_2^2)\omega\tau^3\delta_{\nu,k_1}^{0}\delta_{b,k_2}^{0}\,,\nonumber\\
\delta_{b}^{(2)}=&\frac{R_\nu}{16R_\gamma}(k^2+k_1^2-k_2^2)\tau^2\delta_{\nu,k_1}^{0}\delta_{b,k_2}^{0}\,,\nonumber\\
\delta_{\gamma}^{(2)}=&\frac{R_\nu R_b}{R_\gamma}\left(\frac{2}{3}\omega\tau-\frac{1}{4}(\omega\tau)^2\right)\delta_{\nu,k_1}^{0}\delta_{b,k_2}^{0}\,,\nonumber\\
\delta_{\nu}^{(2)}=&R_b\left(-\frac{2}{3}\omega\tau+\frac{1}{4}(\omega\tau)^2\right)\delta_{\nu,k_1}^{0}\delta_{b,k_2}^{0}\,,\\
v_{\gamma b}^{(2)}=&\left(\frac{(k^2+k_1^2-k_2^2)R_b R_\nu(R_\nu-4)}{32R_\gamma^2 k^2}\omega\tau^2+f_v^{\nu b} (k,k_1,k_2)\omega^2\tau^3\right)\delta_{\nu,k_1}^{0}\delta_{b,k_2}^{0}\,,\nonumber\\
v_{\nu}^{(2)}=&R_b\left(\frac{(k^2+k_1^2-k_2^2)}{32k^2}\omega\tau^2-\frac{(k^2+k_1^2-k_2^2)}{96k^2}\omega^2\tau^3\right)\delta_{\nu,k_1}^{0}\delta_{b,k_2}^{0}\,,\nonumber\\
\sigma_\nu^{(2)}=&f_\sigma^{\nu b}(k,k_1,k_2) \omega k^2 \tau^3\delta_{\nu,k_1}^{0}\delta_{b,k_2}^{0}\,,\nonumber\\
\Delta_3^{(2)}=&O(\tau^4)\,,\nonumber
\end{align}
with the following kernels:
\begin{align}
f^{\nu b}_\psi=&-\frac{R_bR_\nu}{1440R\gamma(2R_\nu+15)^2(4R_\nu+15) k^4}\left[3 (1125 + 3750 R_\nu + 620 R_\nu^2 - 4 R_\nu^3) k^4 \right.\nonumber\\
&- 30 (225 - 195 R_\nu - 32 R_\nu^2 + 2 R_\nu^3) (k_1^2 - k_2^2)^2 \nonumber\\
&\left.+ k^2 ((14625 + 2700 R_\nu - 860 R_\nu^2 + 8 R_\nu^3) k_1^2 + 3 (-1875 - 3500 R_\nu - 140 R_\nu^2 + 24 R_\nu^3) k_2^2)\right]\,,\nonumber\\
f^{\nu b}_E=&\frac{R_b}{R_c}f^{\nu c}_E\,,\nonumber\\
f^{\nu b}_v=&\frac{R_\nu R_b}{384R_\gamma^3 k^2}\left[k^2(-5+R_\nu+4 R_\nu^2+9R_b(5+R_\nu))\right.\nonumber\\
&\left.+(k_1^2-k_2^2)(-5 + R_b (69 - 15 R_\nu) + R_\nu + 4 R_\nu^2)\right]\,,\nonumber\\
f^{\nu b}_\sigma=&\frac{R_b}{R_c}f^{\nu c}_\sigma\,.\nonumber
\end{align}

We can also analyse here if the compensated isocurvature generates an extra contribution when mixed with the neutrino isocurvature. We find that it does and present below the initial evolution for that mixed mode, showing only the non-zero variables:
\begin{align}
\psi^{(2)}=&\frac{R_\nu R_c}{480 R_\gamma}(k^2+k_1^2-k_2^2)\omega\tau^3\delta_{\nu,k_1}^{0}\delta_{CI,k_2}^{0}\,,\nonumber\\
\delta_{c}^{(2)}=&\frac{R_\nu}{160 R_\gamma}\left(R_c(k^2+k_1^2-k_2^2)-2R_b k_1^2\right)\omega\tau^3\delta_{\nu,k_1}^{0}\delta_{CI,k_2}^{0}\,,\nonumber\\
\delta_{b}^{(2)}=&-\frac{R_\nu R_c}{16 R_\gamma R_b}(k^2+k_1^2-k_2^2)\tau^2\delta_{\nu,k_1}^{0}\delta_{CI,k_2}^{0}\,,\nonumber\\
\delta_{\gamma}^{(2)}=&\frac{R_\nu R_c(6-R_\nu)}{120 R_\gamma^2}(k^2+k_1^2-k_2^2)\omega\tau^3\delta_{\nu,k_1}^{0}\delta_{CI,k_2}^{0}\,,\nonumber\\
\delta_{\nu}^{(2)}=&\frac{R_\nu R_c}{120 R_\gamma}(k^2+k_1^2-k_2^2)\omega\tau^3\delta_{\nu,k_1}^{0}\delta_{CI,k_2}^{0}\,,\\
v_{\gamma b}^{(2)}=&\frac{3 R_\nu R_c}{32 R_\gamma^2}\frac{k^2+k_1^2-k_2^2}{k^2}\omega\tau^2\delta_{\nu,k_1}^{0}\delta_{CI,k_2}^{0}\,,\nonumber\\
v_{\nu}^{(2)}=&-\frac{R_\nu R_c}{1920 R_\gamma}(k^2+k_1^2-k_2^2)\omega\tau^4\delta_{\nu,k_1}^{0}\delta_{CI,k_2}^{0}\,.\nonumber
\end{align}
We see that the mixture of these two modes is far more consequential in this case than it was when the compensated isocurvature mixed with the adiabatic mode. In particular, the compensated isocurvature relation, Eq.~\eqref{CIP}, is not conserved at second order and many other quantities are generated besides the matter density perturbations, in clear contrast to what happens at the linear level.

We also note that in all the modes above, the hierarchy between $v_\nu$, $\sigma_\nu$ and $\Delta_3$ is maintained, i.e. $v_\nu\gtrsim\sigma_\nu\gtrsim\Delta_3$, in terms of their order in the expansion in $\tau$. This gives us confidence that we can neglect the initial evolution of the higher brightness tensors for all the modes under study.

\section{Conclusion}\label{conclusion}

We have studied the approximate initial solutions for the transfer functions of the most relevant variables used in the initialization of Boltzmann solvers at second order in perturbation theory. In order to do this, we have described the differential system and precisely defined the different modes under study. We have concluded that the number of purely growing modes is smaller at second order, as we have shown that the neutrino velocity mode sources decaying solutions due to its contribution to the total anisotropic stress. Furthermore, we have highlighted the importance of the solutions sourced by multiple modes, which have no first order counter-part. We show that these solutions exhibit growing behaviour, thus making them essential for the accurate evolution of the cosmological variables.

We also investigated in detail the consequences of a compensated isocurvature mode, the mode which is constrained the least at the linear level. We confirm that a pure compensated isocurvature mode does not generate any evolution both at first and second order in cosmological perturbations. However, we show that, when mixed with other modes, there are additional contributions to many variables, which do not exist at linear order or in the pure compensated mode. In particular, we noted that the mixed adiabatic and compensated isocurvature solution conserves the relation between the baryon and dark matter contrasts given initially, but also causes the compensated density fluctuation to grow, as well as the baryon-photon velocity. Considering the other possible mixture, with the neutrino density isocurvature, we find that the curvature perturbation, density contrasts and velocity perturbations receive a contribution from this mixed mode, but no higher multipoles are affected.

Our results can be applied to initialize second order Boltzmann codes to evaluate the effects of isocurvatures on a variety of observables. In the future, we aim to apply the same techniques developed here to study the initialization of vector modes, which are known to be sourced when multiple degrees of freedom are present. This would be an interesting application for the mixed modes found in this work.

\section*{Acknowledgements}
PC is supported by a Queen Mary Principal's Research Studentship, by a Bolsa de Excel\^encia Acad\'{e}mica of the Funda\c{c}\~{a}o Eug\'{e}nio de Almeida and by the Funda\c{c}\~{a}o para a Ci\^{e}ncia e Tecnologia (FCT) grant SFRH/BD/118740/2016, KAM is
supported, in part, by STFC grant ST/M001202/1.
The tensor algebra package xAct \cite{xAct}\footnote{\href{http://www.xact.es}{http://www.xact.es}}, as well as its
sub-package xPand \cite{xPand}\footnote{\href{http://www.xact.es/xPand/}{http://www.xact.es/xPand/}}, were used in the derivation of
many of the equations presented in this work.
The authors are grateful to David Mulryne and Alkistis Portsidou for useful discussions.

\appendix

\section{Gauge transformations to Poisson gauge}\label{gaugetr}

In order to apply the results of this paper in other settings in which the synchronous gauge is not used, one has to perform a gauge transformation from the synchronous gauge to the desired gauge. We present the general transformations here, as well as the specific transformations to the Poisson gauge applied to each mode.

We begin by defining a gauge generator, $\xi^\mu$, to parametrise the gauge transformation. The effect of a gauge transformation on a tensor field $T$ is given by (\cite{Bruni:1996im,Abramo:1997hu,Malik:2008im,Malik:2012dr})
\be
\widetilde T=e^{\L_\xi}T\,,
\ee
in which $\L_\xi$ is  the Lie derivative in the direction of $\xi$. Expanding the relation above order by order, one finds, up to second order,
\bea
\widetilde{\delta T_1}=\delta T_1+\L_{\xi_1}T_0\,,\\
\widetilde{\delta T_2}=\delta T_2+\L_{\xi_2}T_0+\L_{\xi_1}^2T_0
+2\L_{\xi_1}\delta T_1\,,
\eea
where we have expanded the gauge generator order by order as
$\xi^\mu=\xi_1^\mu+\frac12 \xi_2^\mu+\dots$. We decompose it further into scalar and vector parts as
\be
\xi^\mu=\lb\alpha ,\beta^{,i}+\gamma^i\rb\,.
\ee

With these definitions and the metric defined in the main text in Eqs.~\eqref{metric1}-\eqref{metric3}, the first order gauge transformations of the metric variables are given by
\begin{align}
&\wt{\phi}_1= \phi_1+\Hh\alpha_1+\alpha_1'\,,\ \ \ \ \ \label{gauge1}
\wt{\psi}_1=\psi_1-\Hh\alpha_1\,,\\
&\wt{E}_1= E_1+\beta_1\,,\ \ \ \ \ \ \ \quad\quad\>\wt{B}_1=B_1-\alpha_1+\beta_1'\,,\\
&\wt{F}_1^i= F_1^i+\gamma_1^i\,,\ \ \ \ \ \ \ \quad\quad\ \wt{S}_1^i=S_1^i-\gamma_1^{i\prime}\,,\\
&\wt{h}_1^{ij}= h_1^{ij}\,,
\end{align}
and those of the fluid quantities are
\begin{align}
&\wt{\delta_1}=\delta_1-3\Hh(1+w)\alpha_1\,,\\
&\wt{v_1}=v_1-\beta_1'\,,\ \ \ \ \ \quad\quad\wt{v_{V1}^i}=v_{V1}^i-\gamma_{1}^{i\prime}\,,\label{gauge2}
\end{align}
which are valid for any of the species presented above.

The synchronous gauge, used in the main text, is defined via
\begin{equation}
\wt{\phi}=\wt{B}=0\,,\ \ \ \ \ \wt{S^i}=0\,,
\end{equation}
which means that to apply a gauge transformation from a general gauge to synchronous gauge, the appropriate gauge generators are given by
\begin{align}
&\alpha_1=-\frac1a \left(\int{a \phi_1 \text{d}\tau}-C_\alpha(x^i)\right)\,,\\
&\beta_1=\int{(\alpha_1-B_1\text{d}\tau)}+C_\beta(x^i)\,,\\
&\gamma_1^{i}=\int{S_1^i\text{d}\tau}+C_\gamma^i(x^i)\,.
\end{align}
The constant functions $C_\beta$ and $C_\gamma^i$ can be fixed by a choice of coordinates at the initial hypersurface. The function $C_\alpha$, however, represents a residual gauge freedom that exists in this gauge and can be unambiguously chosen by setting the initial dark matter velocity perturbation to zero, as is done throughout this paper.

The Poisson gauge, to which we want to convert our results in this appendix, is specified by the following choices
\begin{equation}
\wt{E}=\wt{B}=0\,,\ \ \ \ \ \wt{F^i}=0\,,
\end{equation}
which implies that the gauge generator components are, at first order,
\begin{align}
&\alpha_1=B_1-E_1'\,,\\
&\beta_1=-E_1\,,\\
&\gamma_1^i=-F_1^i\,.
\end{align}
In this work, we are interested in a transformation from synchronous to Poisson gauge, thus we may simply re-write the first equation above as $\alpha_1^{\text{S2P}}=-E_1^{\text{S}\prime}$. Therefore the gauge transformations for the scalars depend only on the metric potential $E_1$. For that reason, the difference between variables on both gauges depends on the size of $E$ in each mode, in orders of $\tau$. For example, in the CDM isocurvature mode shown in the main text in Eq.~\eqref{matteriso1}, the metric potential $E$ is $O(\tau^3)$. However, it enters $\alpha$ with a time derivative and is usually multiplied by $\Hh$, thus the gauge transformation will make a difference of order $O(\tau)$ in most variables. At leading order, the CDM isocurvature mode is now given in Poisson gauge by
\begin{align}
\psi=&-\frac{R_c(4R_\nu+15)}{8(15+2R_\nu)}\omega\tau\delta_c^0\,,\nonumber\\
\phi=&\frac{R_c(4R_\nu-15)}{8(15+2R_\nu)}\omega\tau\delta_c^0\,,\nonumber\\
\delta_{c}=&\left(1-\frac{3R_c(4R_\nu+15)}{8(15+2R_\nu)}\omega\tau\right)\delta_c^0\,,\nonumber\\
\delta_{b}=&-\frac{3R_c(4R_\nu+15)}{8(15+2R_\nu)}\omega\tau\delta_c^0\,,\nonumber\\
\delta_{\gamma}=&-\frac{R_c(4R_\nu+15)}{2(15+2R_\nu)}\omega\tau\delta_c^0\,,\nonumber\\
\delta_{\nu}=&-\frac{R_c(4R_\nu+15)}{2(15+2R_\nu)}\omega\tau\delta_c^0\,,\\
v_c=&\frac{R_c(15-4R_\nu)}{24(15+2R_\nu)}\omega\tau^2\delta_c^0\,,\nonumber\\
v_{\gamma b}=&\frac{15 R_c}{8(15+2R_\nu)}\omega\tau^2\delta_c^0\,,\nonumber\\
v_{\nu}=&\frac{15R_c}{8(15+2R_\nu)}\omega\tau^2\delta_c^0\,,\nonumber\\
\sigma_\nu=&-\frac{R_c}{6(15+2 R_\nu)}k^2\omega\tau^3\delta_c^0\,.\nonumber
\end{align}

In other modes, the transformation is similar, but can introduce additional issues. For example, in the case of the neutrino velocity isocurvature, some variables will have decaying solutions already at linear order, as $E$ is $O(\tau)$ in that case. This is described, for example, in Ref.~\cite{Shaw:2009nf}, in which the potentials $\phi$ and $\psi$ are given in Poisson gauge for all five linear growing modes. We do not comment further on this issue, as we do not study the neutrino velocity mode at second order, for the reasons explained in the main text.

At second order, the transformation rules become more complex, but can be similarly constructed. They can be consulted in Ref.~\cite{Malik:2008im} and, for brevity, we shall not reproduce them here. In practice, they are very similar to Eqs.~\eqref{gauge1}--\eqref{gauge2}, with the addition of non-linear terms. As before, one can then calculate the form of the gauge generators required to transform from synchronous gauge to Poisson gauge. Applying those transformations to the results in the main text, we find the results for Poisson gauge, which we show in the same order as before, starting with the adiabatic sourced mode. Note, however, that the defining variables (e.g. $\psi^{0}_{k_1}\psi^{0}_{k_2}$) still refer to those variables in synchronous gauge.

\subsection{Pure adiabatic mode}

\begin{align}
\psi=&f^{\psi\psi}_{\psi,P}\psi^{0}_{k_1}\psi^{0}_{k_2}\,,\nonumber\\
\phi=&\left(\frac{20(35+8 R_\nu)}{(15+4R_\nu)^2}-2 f^{\psi\psi}_{\psi,P}\right)\psi^{0}_{k_1}\psi^{0}_{k_2}\,,\nonumber\\
\delta_{c}=&\delta_{b}=\left(-\frac{15(35+16 R_\nu)}{(15+4R_\nu)^2}+3 f^{\psi\psi}_{\psi,P}\right)\psi^{0}_{k_1}\psi^{0}_{k_2}\,,\nonumber\\
\delta_{\gamma}=&\delta_{\nu}=\left(-\frac{40(15+8 R_\nu)}{(15+4R_\nu)^2}+4 f^{\psi\psi}_{\psi,P}\right)\psi^{0}_{k_1}\psi^{0}_{k_2}\,,\\
v_c=&v_{\gamma b}=v_{\nu}=\left(-\frac{40(10+3 R_\nu)}{(15+4R_\nu)^2}+f^{\psi\psi}_{\psi,P}\right)\tau\psi^{0}_{k_1}\psi^{0}_{k_2}\,,\nonumber\\
\sigma_\nu=&-\frac{9 k^4 - 3 (k_1^2 - k_2^2)^2 + 2 k^2 (k_1^2+k_2^2)}{3(15+4R_\nu)k^4}(k\tau)^2\psi^{0}_{k_1}\psi^{0}_{k_2}\,,\nonumber
\end{align}
with,
\begin{align}
f^{\psi\psi}_{\psi,P}=&\frac{5}{(15 + 4 R_\nu)^2 k^4} \left[(25 + 9 R_\nu) k^4 -(5+R_\nu)\left(3 (k_1^2 - k_2^2)^2 - 2 k^2 (k_1^2+k_2^2)\right)\right]\,.\nonumber
\end{align}

We can very easily verify that the adiabatic condition at second order, given in Eq.~\eqref{Adconds2}, is indeed verified in this gauge, as it must. Furthermore, we can now directly compare these results to those given in Refs.~\cite{Pettinari:2014vja,Pitrou:2010sn}. They do not exactly match, due to a different choice of defining variable --- we choose $\psi=-\zeta$, while they choose $\zeta_D=\zeta+\zeta^2$, as defined, for example, in Ref.~\cite{Carrilho:2015cma}. Applying this transformation to the general solution in terms of transfer functions, we find
\begin{align}
X(\tau,k)&=\mathcal{T}^{(1)} \psi^0(k)+\frac12\int_{k_1,k_2}\mathcal{T}^{(2)} \psi^0(k_1)\psi^0(k_2)\\
&=-\mathcal{T}^{(1)} \zeta_D^0(k)+\frac12\int_{k_1,k_2}\lb2\mathcal{T}^{(1)}+\mathcal{T}^{(2)}\rb \zeta_D^0(k_1)\zeta_D^0(k_2)\,,
\end{align}
which shows that, in terms of $\zeta_D$, the second order transfer functions receive an extra contribution of twice the linear transfer function. This is exactly the difference we find between our results and those of Refs.~\cite{Pettinari:2014vja,Pitrou:2010sn}, confirming the match between all results. Care must be taken, however, when these results are applied to situations in which one assumes the initial conditions to be Gaussian. In that case, one must make clear which of the variables has that property, since should $\zeta_D$ be Gaussian, $\zeta$ will not be and vice versa.

\subsection{Pure cold dark matter isocurvature mode}

\begin{align}
\psi=&\left(-\frac{5(15 + 4 R_\nu)^2}{ 64 (15 + 2 R_\nu)^2} R_c^2+\frac13f^{cc}_{b,P}\right)(\omega\tau)^2\delta^{0}_{c,k_1}\delta^{0}_{c,k_2}\,,\nonumber\\
\phi=&\left(\frac{8325 + 2280 R_\nu + 272 R_\nu^2}{ 64 (15 + 2 R_\nu)^2} R_c^2-\frac43f^{cc}_{b,P}\right)(\omega\tau)^2\delta^{0}_{c,k_1}\delta^{0}_{c,k_2}\,,\nonumber\\
\delta_{c}=&\left(-\frac{3(15+4R_\nu)}{4(15+2R_\nu)}R_c\omega\tau+\left(\frac{3(675 + 230 R_\nu + 8 R_\nu^2)}{ 16 (15 + 2 R_\nu)(25+2 R_\nu)} R_c+f^{cc}_{b,P}\right)(\omega\tau)^2\right)\delta^{0}_{c,k_1}\delta^{0}_{c,k_2}\,,\nonumber\\
\delta_{b}=&f^{cc}_{b,P}(\omega\tau)^2\delta^{0}_{c,k_1}\delta^{0}_{c,k_2}\,,\nonumber\\
\delta_{\gamma}=&\delta_{\nu}=\left(\frac{(15+4R_\nu)^2}{ 16 (15 + 2 R_\nu)^2} R_c^2+\frac43f^{cc}_{b,P}\right)(\omega\tau)^2\delta^{0}_{c,k_1}\delta^{0}_{c,k_2}\,,\\
v_c=&\left(-\frac{5(1305+360R_\nu+32R_\nu^2)}{ 192 (15 + 2 R_\nu)^2} R_c^2+\frac13f^{cc}_{b,P}\right)\omega^2\tau^3\delta^{0}_{c,k_1}\delta^{0}_{c,k_2}\,,\nonumber\\
v_{\gamma b}=&v_{\nu}=\left(-\frac{2925+780R_\nu+64R_\nu^2}{ 64 (15 + 2 R_\nu)^2} R_c^2+\frac13f^{cc}_{b,P}\right)\omega^2\tau^3\delta^{0}_{c,k_1}\delta^{0}_{c,k_2}\,,\nonumber\\
\sigma_\nu=&f^{cc}_{\sigma,P}R_c^2\omega^2k^2\tau^4\delta^{0}_{c,k_1}\delta^{0}_{c,k_2}\,,\nonumber
\end{align}
with,
\begin{align}
f^{cc}_{b,P}=&\frac{3R_c^2}{128 (25 + 2 R_\nu) (15 + 2 R_\nu)^2 k^4} \left[(88875 + 42150 R_\nu + 6160 R_\nu^2 + 256 R_\nu^3) k^4 \right.\nonumber\\
&\left.+ 15(-225 + 110 R_\nu + 16 R_\nu^2)\left(3 (k_1^2 - k_2^2)^2 - 2 k^2 (k_1^2 + k_2^2)\right)\right]\,,\nonumber\\
f^{cc}_{\sigma,P}=&-\frac{5(855+138R_\nu+4R_\nu^2)k^4+(825+70R_\nu-4R_\nu^2)\left(3(k_1^2-k_2^2)^2-2k^2(k_1^2+k_2^2)\right)}{48(15+2R_\nu)^2(25+2R_\nu)}\,.\nonumber
\end{align}

\subsection{Mixture of adiabatic and cold dark matter modes}

\begin{align}
\psi=&f^{c\psi}_{\psi,P}\omega\tau\delta^{0}_{c,k_1}\psi^{0}_{k_2}\,,\nonumber\\
\phi=&\left(\frac{75+8 R_\nu(20+3 R_\nu)}{2(15+2R_\nu)(15+4R_\nu)}R_c-3f^{c\psi}_{\psi,P}\right)\omega\tau\delta^{0}_{c,k_1}\psi^{0}_{k_2}\,,\nonumber\\
\delta_{c}=&\left(-\frac{15}{15+4R_\nu}+\left(-\frac{3\left(75(1+R_c)+4 R_\nu(20+(35+8 R_\nu)R_c)\right)}{8(15+2R_\nu)(15+4R_\nu)}+3f^{c\psi}_{\psi,P}\right)\omega\tau\right)\delta^{0}_{c,k_1}\psi^{0}_{k_2}\,,\nonumber\\
\delta_{b}=&\left(-\frac{3(5+8 R_\nu)}{8(15+2R_\nu)}R_c+3f^{c\psi}_{\psi,P}\right)\omega\tau\delta^{0}_{c,k_1}\psi^{0}_{k_2}\,,\nonumber\\
\delta_{\gamma}=&\delta_{\nu}=\left(-\frac{4 R_\nu}{15+2R_\nu}R_c+4f^{c\psi}_{\psi,P}\right)\omega\tau\delta^{0}_{c,k_1}\psi^{0}_{k_2}\,,\\
v_c=&\left(-\frac{(35+8R_\nu)(k^2+k_1^2-k_2^2)}{24(15+4R_\nu)k^2}+f^{c\psi}_{v,P}\right)\omega\tau^2\delta^{0}_{c,k_1}\psi^{0}_{k_2}\,,\nonumber\\
v_{\gamma b}=&v_{\nu}=f^{c\psi}_{v,P}\omega\tau^2\delta^{0}_{c,k_1}\psi^{0}_{k_2}\,,\nonumber\\
\sigma_\nu=&f^{c\psi}_{\sigma,P}\omega k^2\tau^3\delta^{0}_{c,k_1}\psi^{0}_{k_2}\,,\nonumber
\end{align}
with,
\begin{align}
f^{c\psi}_{\psi,P}=&\frac{R_c}{16 (15 + 2 R_\nu) (15 + 4 R_\nu) k^4} \left[(375 + 315 R_\nu + 64 R_\nu^2) k^4\right.\nonumber\\
&\left. - 45 (-5 + R_\nu) (k_1^2 - k_2^2)^2 + 30 k^2 ((-5 + 3 R_\nu) k_1^2 - (5 + R_\nu) k_2^2)\right]\,,\nonumber
\end{align}
\begin{align}
f^{c\psi}_{v,P}=&\frac{5R_c}{16 (15 + 2 R_\nu) (15 + 4 R_\nu) k^4} \left[(135 + 19 R_\nu) k^4 \right.\nonumber\\
&\left.- 9 (-5 + R_\nu) (k_1^2 - k_2^2)^2 + k^2 ((30 + 38 R_\nu) k_1^2 - 2 (45 + 13 R_\nu) k_2^2)\right]\,,\nonumber\\
f^{c\psi}_{\sigma,P}=&\frac{R_c}{12 (15 + 2 R_\nu) (15 + 4 R_\nu) k^4} \left[(75 + 4 R_\nu) k^4 \right.\nonumber\\
&\left.- 3 (-5 + 4 R_\nu) (k_1^2 - k_2^2)^2 + k^2 ((-70 + 8 R_\nu) k_1^2 + 2 (25 + 4 R_\nu) k_2^2)\right]\,.\nonumber
\end{align}

\subsection{Pure baryon isocurvature mode}

\begin{align}
\psi=&\left(-\frac{5(15+ + 4 R_\nu)^2}{ 64 (15 + 2 R_\nu)^2} R_b^2+\frac13f^{bb}_{c,P}\right)(\omega\tau)^2\delta^{0}_{b,k_1}\delta^{0}_{b,k_2}\,,\nonumber\\
\phi=&\left(\frac{8325 + 2280 R_\nu + 272 R_\nu^2}{ 64 (15 + 2 R_\nu)^2} R_b^2-\frac43f^{bb}_{c,P}\right)(\omega\tau)^2\delta^{0}_{b,k_1}\delta^{0}_{b,k_2}\,,\nonumber\\
\delta_{c}=&f^{bb}_{c,P}(\omega\tau)^2\delta^{0}_{b,k_1}\delta^{0}_{b,k_2}\,,\nonumber\\
\delta_{b}=&\left(-\frac{3(15+4R_\nu)}{4(15+2R_\nu)}R_b\omega\tau+\left(\frac{3(675 + 230 R_\nu + 8 R_\nu^2)}{ 16 (15 + 2 R_\nu)(25+2 R_\nu)} R_b+f^{bb}_{c,P}\right)(\omega\tau)^2\right)\delta^{0}_{b,k_1}\delta^{0}_{b,k_2}\,,\nonumber\\
\delta_{\gamma}=&\delta_{\nu}=\left(\frac{(15+4R_\nu)^2}{ 16 (15 + 2 R_\nu)^2} R_b^2+\frac43f^{bb}_{c,P}\right)(\omega\tau)^2\delta^{0}_{b,k_1}\delta^{0}_{b,k_2}\,,\\
v_c=&\left(-\frac{5(1305+360R_\nu+32R_\nu^2)}{ 192 (15 + 2 R_\nu)^2} R_b^2+\frac13f^{bb}_{c,P}\right)\omega^2\tau^3\delta^{0}_{b,k_1}\delta^{0}_{b,k_2}\,,\nonumber\\
v_{\gamma b}=&\left(-\frac{3825-1905R_\nu-700R_\nu^2-64R_\nu^3}{ 64 R_\gamma(15 + 2 R_\nu)^2} R_b^2+\frac13f^{bb}_{c,P}\right)\omega^2\tau^3\delta^{0}_{b,k_1}\delta^{0}_{b,k_2}\,,\nonumber\\
v_{\nu}=&\left(-\frac{2925+780R_\nu+64R_\nu^2}{ 64 (15 + 2 R_\nu)^2} R_b^2+\frac13f^{bb}_{c,P}\right)\omega^2\tau^3\delta^{0}_{b,k_1}\delta^{0}_{b,k_2}\,,\nonumber\\
\sigma_\nu=&f^{cc}_{\sigma,P}R_b^2\omega^2k^2\tau^4\delta^{0}_{b,k_1}\delta^{0}_{b,k_2}\,,\nonumber
\end{align}
with,
\begin{align}
f^{bb}_{c,P}=&\frac{R_b^2}{R_c^2}f^{cc}_{b,P}\,.\nonumber
\end{align}

\subsection{Mixture of baryon and cold dark matter modes}

\begin{align}
\psi=&f^{bc}_{\psi,P}(\omega\tau)^2\delta^{0}_{b,k_1}\delta^{0}_{c,k_2}\,,\nonumber\\
\phi=&\left(\frac{3(1275 - 40 R_\nu - 16 R_\nu^2}{ 64 (15 + 2 R_\nu)^2} R_bR_c-4f^{bc}_{\psi,P}\right)(\omega\tau)^2\delta^{0}_{b,k_1}\delta^{0}_{c,k_2}\,,\nonumber\\
\delta_{c}=&\left[-3R_b\frac{15+4R_\nu}{15+2R_\nu}\omega\tau+\left(f^{bc}_\delta(R_b)+3f^{bc}_{\psi,P}\right)(\omega\tau)^2\right]\delta^{0}_{b,k_1}\delta^{0}_{c,k_2}\,,\nonumber\\
\delta_{b}=&\left[-3R_c\frac{15+4R_\nu}{15+2R_\nu}\omega\tau+\left(f^{bc}_\delta(R_c)+3f^{bc}_{\psi,P}\right)(\omega\tau)^2\right]\delta^{0}_{b,k_1}\delta^{0}_{c,k_2}\,,\nonumber\\
\delta_{\gamma}=&\delta_{\nu}=\left(-\frac{3(15+4R_\nu)^2}{ 8 (15 + 2 R_\nu)^2} R_bR_c+4f^{bc}_{\psi,P}\right)(\omega\tau)^2\delta^{0}_{b,k_1}\delta^{0}_{c,k_2}\,,\\
v_c=&\left(-\frac{5(315-8R_\nu^2)}{ 96(15 + 2 R_\nu)^2} R_bR_c+f^{bc}_{\psi,P}\right)\omega^2\tau^3\delta^{0}_{b,k_1}\delta^{0}_{c,k_2}\,,\nonumber\\
v_{\gamma b}=&\left(-\frac{(1125-750R_\nu-94R_\nu^2+8R_\nu^3)k^2-(15+2R_\nu)^2(k_1^2-k_2^2)}{ 32 R_\gamma(15 + 2 R_\nu)^2k^2} R_bR_c+f^{bc}_{\psi,P}\right)\omega^2\tau^3\delta^{0}_{b,k_1}\delta^{0}_{c,k_2}\,,\nonumber\\
v_{\nu}=&\left(-\frac{450+45R_\nu-4R_\nu^2}{ 16 (15 + 2 R_\nu)^2} R_bR_c+f^{bc}_{\psi,P}\right)\omega^2\tau^3\delta^{0}_{b,k_1}\delta^{0}_{c,k_2}\,,\nonumber\\
\sigma_\nu=&f^{bc}_{\sigma,P}R_b^2\omega^2k^2\tau^4\delta^{0}_{b,k_1}\delta^{0}_{c,k_2}\,,\nonumber
\end{align}
with,
\begin{align}
f^{bc}_{\psi,P}=&-\frac{R_bR_c}{128(25+2R_\nu)(15+2R_\nu)^2k^4}\left[(-32625 - 7650 R_\nu + 240 R_\nu^2 + 64 R_\nu^3) k^4 \right.\nonumber\\\nonumber
&\left.- 15 (-225 + 110 R_\nu + 16 R_\nu^2)\left(3 (k_1^2 - k_2^2)^2 -2 k^2 (k_1^2 + k_2^2)\right)\right]\,,\\
f^{bc}_\delta(R_x)=&\frac{3R_x\left(48375-5R_x(25+2R_\nu)(15+4R_\nu)^2+2R_\nu(13425+4R_\nu(545+24R_\nu))\right)}{64(25+2R_\nu)(15+2R_\nu)^2}\,,\nonumber\\
f^{bc}_{\sigma,P}=&\frac{R_bR_c}{48(25+2R_\nu)(15+2R_\nu)^2k^4}\left[5 (855 + 138 R_\nu + 4 R_\nu^2) k^4 \right.\nonumber\\\nonumber
&\left.- (-825 - 70 R_\nu + 4 R_\nu^2)\left(3 (k_1^2 - k_2^2)^2 - 2 k^2 (k_1^2 + k_2^2)\right)\right]\,.
\end{align}

\subsection{Mixture of adiabatic and baryon modes}

\begin{align}
\psi=&f^{b\psi}_{\psi,P}\omega\tau\delta^{0}_{b,k_1}\psi^{0}_{k_2}\,,\nonumber\\
\phi=&\left(\frac{75+8 R_\nu(20+3 R_\nu)}{2(15+2R_\nu)(15+4R_\nu)}R_b-3f^{c\psi}_{\psi,P}\right)\omega\tau\delta^{0}_{b,k_1}\psi^{0}_{k_2}\,,\nonumber\\
\delta_{c}=&\left(-\frac{3(5+8 R_\nu)}{8(15+2R_\nu)}R_b+3f^{b\psi}_{\psi,P}\right)\omega\tau\delta^{0}_{b,k_1}\psi^{0}_{k_2}\,,\nonumber\\
\delta_{b}=&\left(-\frac{15}{15+4R_\nu}+\left(-\frac{3\left(75(1+R_c)+4 R_\nu(20+(35+8 R_\nu)R_b)\right)}{8(15+2R_\nu)(15+4R_\nu)}+3f^{b\psi}_{\psi,P}\right)\omega\tau\right)\delta^{0}_{b,k_1}\psi^{0}_{k_2}\,,\nonumber\\
\delta_{\gamma}=&\delta_{\nu}=\left(-\frac{4 R_\nu}{15+2R_\nu}R_c+4f^{b\psi}_{\psi,P}\right)\omega\tau\delta^{0}_{b,k_1}\psi^{0}_{k_2}\,,\\
v_c=&\left(-\frac{(35+8R_\nu)(k^2+k_1^2-k_2^2)}{24(15+4R_\nu)k^2}R_b+f^{b\psi}_{v,P}\right)\omega\tau^2\delta^{0}_{b,k_1}\psi^{0}_{k_2}\,,\nonumber\\
v_{\gamma b}=&v_{\nu}=f^{b\psi}_{v,P}\omega\tau^2\delta^{0}_{b,k_1}\psi^{0}_{k_2}\,,\nonumber\\
\sigma_\nu=&f^{b\psi}_{\sigma,P}\omega k^2\tau^3\delta^{0}_{b,k_1}\psi^{0}_{k_2}\,,\nonumber
\end{align}
with,
\begin{align}
f^{b\psi}_{\psi,P}=&\frac{R_b}{R_c}f^{c\psi}_{\psi,P}\,,\nonumber\\
f^{b\psi}_{v,P}=&\frac{R_b}{R_c}f^{c\psi}_{v,P}\,,\nonumber\\
f^{b\psi}_{\sigma,P}=&\frac{R_b}{R_c}f^{c\psi}_{\sigma,P}\,.\nonumber
\end{align}

\subsection{Mixture of adiabatic and compensated modes}

\begin{align}
\delta_{c}=&-\frac{R_b}{R_c}\delta_{b}=\left(-\frac{15}{15+4R_\nu}-\frac{15\left(15+16 R_\nu\right)}{8(15+2R_\nu)(15+4R_\nu)}\omega\tau\right)\delta^{0}_{CI,k_1}\psi^{0}_{k_2}\,,\nonumber\\
v_{\gamma b}=&\frac{\left(k^2-k_1^2+k_2^2\right)R_c}{96R_\gamma k^2}k_2^2\omega\tau^4 \delta^{0}_{CI,k_1}\psi^{0}_{k_2}\,.
\end{align}

\subsection{Pure neutrino density isocurvature mode}

\begin{align}
\psi=&f^{\nu\nu}_{\psi,P}\delta^{0}_{\nu,k_1}\delta^{0}_{\nu,k_2}\,,\nonumber\\
\phi=&\left(\frac{4R_\nu^2}{(15+4R_\nu)^2}-2 f^{\nu\nu}_{\psi,P}\right)\delta^{0}_{\nu,k_1}\delta^{0}_{\nu,k_2}\,,\nonumber\\
\delta_{c}=&\delta_{b}=\left(\frac{15R_\nu^2}{(15+4R_\nu)^2}+3 f^{\nu\nu}_{\psi,P}\right)\delta^{0}_{\nu,k_1}\delta^{0}_{\nu,k_2}\,,\nonumber\\
\delta_{\gamma}=&\left(-\frac{8R_\nu(12+7R_\nu)}{R_\gamma(15+4R_\nu)^2}+4 f^{\nu\nu}_{\psi,P}\right)\delta^{0}_{\nu,k_1}\delta^{0}_{\nu,k_2}\,,\nonumber\\
\delta_{\nu}=&\left(\frac{8R_\nu(15+7R_\nu)}{(15+4R_\nu)^2}+4 f^{\nu\nu}_{\psi,P}\right)\delta^{0}_{\nu,k_1}\delta^{0}_{\nu,k_2}\,,\\
v_c=&\left(\frac{2R_\nu^2}{(15+4R_\nu)^2}+f^{\nu\nu}_{\psi,P}\right)\tau\delta^{0}_{\nu,k_1}\delta^{0}_{\nu,k_2}\,,\nonumber\\
v_{\gamma b}=&\left(\frac{R_\nu^2(233+8R_\nu(13+3R_\nu))}{4R_\gamma^2(15+4R_\nu)^2}+f^{\nu\nu}_{\psi,P}\right)\tau\delta^{0}_{\nu,k_1}\delta^{0}_{\nu,k_2}\,,\nonumber\\
v_{\nu}=&\left(\frac{3(75+8R_\nu(5+R_\nu))}{4(15+4R_\nu)^2}+f^{\nu\nu}_{\psi,P}\right)\tau\delta^{0}_{\nu,k_1}\delta^{0}_{\nu,k_2}\,,\nonumber\\
\sigma_\nu=&f^{\nu\nu}_{\sigma,P}(k\tau)^2\delta^{0}_{\nu,k_1}\delta^{0}_{\nu,k_2}\,,\nonumber
\end{align}
with,
\begin{align}
f^{\nu\nu}_{\psi,P}=&-\frac{R_\nu^2\left((1-96R_\nu)k^4+285(k_1^2-k_2^2)^2-190k^2(k_1^2+k_2^2)\right)}{16  R_\gamma (15 + 4 R_\nu)^2 k^4} \,,\nonumber\\
f^{\nu\nu}_{\sigma,P}=&\frac{1}{48 R_\gamma (15 + 4 R_\nu)^2 k^4} \left[(-225 - 39 R_\nu + 188 R_\nu^2) k^4 \right.\nonumber\\
&\left.+  (225 - 153 R_\nu + 4 R_\nu^2)\left(3 (k_1^2 - k_2^2)^2 - 2 k^2 (k_1^2 + k_2^2)\right)\right]\,.\nonumber
\end{align}

\subsection{Mixture of adiabatic and neutrino modes}

\begin{align}
\psi=&f^{\nu\psi}_{\psi,P}\delta^{0}_{\nu,k_1}\psi^{0}_{k_2}\,,\nonumber\\
\phi=&\left(-\frac{16R_\nu(5+R_\nu)}{(15+4R_\nu)^2}+2 f^{\nu\psi}_{\psi,P}\right)\delta^{0}_{\nu,k_1}\psi^{0}_{k_2}\,,\nonumber\\
\delta_{c}=&\delta_{b}=\left(\frac{3R_\nu(5+8R_\nu)}{(15+4R_\nu)^2}+3 f^{\nu\psi}_{\psi,P}\right)\delta^{0}_{\nu,k_1}\psi^{0}_{k_2}\,,\nonumber\\
\delta_{\gamma}=&\left(\frac{4R_\nu(75+4R_\nu(7-2R_\nu))}{R_\gamma(15+4R_\nu)^2}+4 f^{\nu\psi}_{\psi,P}\right)\delta^{0}_{\nu,k_1}\psi^{0}_{k_2}\,,\nonumber\\
\delta_{\nu}=&\left(-\frac{4(75+4R_\nu(5-2R_\nu))}{(15+4R_\nu)^2}+4 f^{\nu\psi}_{\psi,P}\right)\delta^{0}_{\nu,k_1}\psi^{0}_{k_2}\,,\\
v_c=&f^{\nu\psi}_{v,P}\tau\delta^{0}_{\nu,k_1}\psi^{0}_{k_2}\,,\nonumber\\
v_{\gamma b}=&\left(\frac{R_\nu(k^2+k_1^2-k_2^2)}{4R_\gamma k^2}+f^{\nu\psi}_{v,P}\right)\tau\delta^{0}_{\nu,k_1}\psi^{0}_{k_2}\,,\nonumber\\
v_{\nu}=&\left(-\frac{k^2+k_1^2-k_2^2}{4 k^2}+f^{\nu\psi}_{v,P}\right)\tau\delta^{0}_{\nu,k_1}\psi^{0}_{k_2}\,,\nonumber\\
\sigma_\nu=&f^{\nu\psi}_{\sigma,P}(k\tau)^2\delta^{0}_{\nu,k_1}\psi^{0}_{k_2}\,,\nonumber
\end{align}
with,
\begin{align}
f^{\nu\psi}_{\psi,P}=&\frac{R_\nu\left((-55 - 32 R_\nu) k^4 + 45 (k_1^2 - k_2^2)^2 + 10 k^2 (-7 k_1^2 + k_2^2)\right)}{4 (15 + 4 R_\nu)^2 k^4} \,,\nonumber\\
f^{\nu\psi}_{v,P}=&\frac{R_\nu\left((85 + 16 R_\nu) k^4 + 45 (k_1^2 - k_2^2)^2 + 2 k^2 ((-5 + 8 R_\nu) k_1^2 - (25 + 8 R_\nu) k_2^2)\right)}{4 (15 + 4 R_\nu)^2 k^4} \,,\nonumber\\
f^{\nu\psi}_{\sigma,P}=&-\frac{1}{4 (15 + 4 R_\nu)^2 k^4} \left[(45 + 4 R_\nu) k^4 - 3 (5 + 4 R_\nu) (k_1^2 - k_2^2)^2 \right.\nonumber\\
&\left. + k^2 ((-30 + 8 R_\nu) k_1^2 + 2 (25 + 4 R_\nu) k_2^2)\right]\,.\nonumber
\end{align}

\subsection{Mixture of dark matter and neutrino modes}

\begin{align}
\psi=&f^{\nu c}_{\psi,P}\omega\tau\delta^{0}_{\nu,k_1}\delta^{0}_{c,k_2}\,,\nonumber\\
\phi=&\left(\frac{R_\nu R_c(105+8R_\nu)}{4(15+2R_\nu)(15+4R_\nu)}-3 f^{\nu c}_{\psi,P}\right)\omega\tau\delta^{0}_{\nu,k_1}\delta^{0}_{c,k_2}\,,\nonumber\\
\delta_{c}=&\left(\frac{3R_\nu}{15+4R_\nu}+\left(\frac{3R_\nu(5R_b(15+4R_\nu)-105-16R_\nu)}{8(15+2R_\nu)(15+4R_\nu)}+3 f^{\nu c}_{\psi,P}\right)\omega\tau\right)\delta^{0}_{\nu,k_1}\delta^{0}_{c,k_2}\,,\nonumber\\
\delta_{b}=&\left(-\frac{15R_\nu R_c}{8(15+2R_\nu)}+3 f^{\nu c}_{\psi,P}\right)\omega\tau\delta^{0}_{\nu,k_1}\delta^{0}_{c,k_2}\,,\nonumber\\
\delta_{\gamma}=&\left(\frac{R_\nu R_c(9+10R_\nu)}{2 R_\gamma (15+2R_\nu)}+4 f^{\nu c}_{\psi,P}\right)\omega\tau\delta^{0}_{\nu,k_1}\delta^{0}_{c,k_2}\,,\nonumber\\
\delta_{\nu}=&\left(-\frac{5R_c(3+2R_\nu)}{2(15+2R_\nu)}+4 f^{\nu c}_{\psi,P}\right)\omega\tau\delta^{0}_{\nu,k_1}\delta^{0}_{c,k_2}\,,\\
v_c=&\left(-\frac{R_c\left((45+8R_\nu)k^2+(45+16R_\nu)(k_1^2-k_2^2)\right)}{96(15+4R_\nu)k^2}+f^{\nu c}_{v,P}\right)\omega\tau^2\delta^{0}_{\nu,k_1}\delta^{0}_{c,k_2}\,,\nonumber\\
v_{\gamma b}=&\left(-\frac{R_c(k^2+k_1^2-k_2^2)}{32R_\gamma k^2}+f^{\nu c}_{v,P}\right)\omega\tau^2\delta^{0}_{\nu,k_1}\delta^{0}_{c,k_2}\,,\nonumber\\
v_{\nu}=&f^{\nu c}_{v,P}\omega\tau^2\delta^{0}_{\nu,k_1}\delta^{0}_{c,k_2}\,,\nonumber\\
\sigma_\nu=&f^{\nu c}_{\sigma,P}(k\tau)^2\delta^{0}_{\nu,k_1}\delta^{0}_{c,k_2}\,,\nonumber
\end{align}
with,
\begin{align}
f^{\nu c}_{\psi,P}=&\frac{3R_\nu R_c}{64 (15 + 2 R_\nu)^2 (15 + 4 R_\nu) k^4} \left[(975 + 550 R_\nu + 64 R_\nu^2) k^4 \right.\nonumber\\
&\left.- 15 (135 + 22 R_\nu) (k_1^2 - k_2^2)^2 + 10 k^2 ((105 + 26 R_\nu) k_1^2 + 3 (55 + 6 R_\nu) k_2^2)\right]\,,\nonumber\\
f^{\nu c}_{v,P}=&-\frac{15 R_c}{64 (15 + 2 R_\nu)^2 (15 + 4 R_\nu) k^4} \left[3 (-150 + 55 R_\nu + 14 R_\nu^2) k^4 \right.\nonumber\\
&+ 3 R_\nu (135 + 22 R_\nu) (k_1^2 - k_2^2)^2 \nonumber\\
&\left.- 2 k^2 (15 (15 + 11 R_\nu + 2 R_\nu^2) k_1^2 + (-225 + 105 R_\nu + 14 R_\nu^2) k_2^2)\right]\,,\nonumber\\
f^{\nu c}_{\sigma,P}=&-\frac{R_c}{48 (15 + 2 R_\nu)^2 (15 + 4 R_\nu) k^4} \left[3 (-1125 - 150 R_\nu + 8 R_\nu^2) (k^4 + (k_1^2 - k_2^2)^2) \right.\nonumber\\\nonumber
&\left.- 2 k^2 ((-675 - 210 R_\nu + 8 R_\nu^2) k_1^2 + (-1575 - 90 R_\nu + 8 R_\nu^2) k_2^2)\right]\,.
\end{align}

\subsection{Mixture of baryon and neutrino modes}

\begin{align}
\psi=&f^{\nu c}_{\psi,P}\omega\tau\delta^{0}_{\nu,k_1}\delta^{0}_{b,k_2}\,,\nonumber\\
\phi=&\left(\frac{R_\nu R_b(105+8R_\nu)}{4(15+2R_\nu)(15+4R_\nu)}-3 f^{\nu b}_{\psi,P}\right)\omega\tau\delta^{0}_{\nu,k_1}\delta^{0}_{b,k_2}\,,\nonumber\\
\delta_{c}=&\left(-\frac{15R_\nu R_b}{8(15+2R_\nu)}+3 f^{\nu b}_{\psi,P}\right)\omega\tau\delta^{0}_{\nu,k_1}\delta^{0}_{b,k_2}\,,\nonumber\\
\delta_{b}=&\left(\frac{3R_\nu}{15+4R_\nu}+\left(-\frac{3R_\nu(5R_b(15+4R_\nu)+30-4R_\nu)}{8(15+2R_\nu)(15+4R_\nu)}+3 f^{\nu b}_{\psi,P}\right)\omega\tau\right)\delta^{0}_{\nu,k_1}\delta^{0}_{b,k_2}\,,\nonumber\\
\delta_{\gamma}=&\left(\frac{R_\nu R_b(9+10R_\nu)}{2 R_\gamma (15+2R_\nu)}+4 f^{\nu b}_{\psi,P}\right)\omega\tau\delta^{0}_{\nu,k_1}\delta^{0}_{b,k_2}\,,\nonumber\\
\delta_{\nu}=&\left(-\frac{5R_b(3+2R_\nu)}{2(15+2R_\nu)}+4 f^{\nu b}_{\psi,P}\right)\omega\tau\delta^{0}_{\nu,k_1}\delta^{0}_{b,k_2}\,,\\
v_c=&\left(-\frac{R_bR_\nu\left(5(21+4R_\nu)k^2+2(15+2R_\nu)(k_1^2-k_2^2)\right)}{12(15+2R_\nu)(15+4R_\nu)k^2}+f^{\nu b}_{\psi,P}\right)\omega\tau^2\delta^{0}_{\nu,k_1}\delta^{0}_{b,k_2}\,,\nonumber\\
v_{\gamma b}=&\left(f^{\nu b}_{v,P}+f^{\nu b}_{\psi,P}\right)\omega\tau^2\delta^{0}_{\nu,k_1}\delta^{0}_{b,k_2}\,,\nonumber\\
v_{\nu}=&\left(-\frac{3R_b\left((2R_\nu(35+8R_\nu)-75)k^2-5(15+2R_\nu)(k_1^2-k_2^2)\right)}{32(15+2R_\nu)(15+4R_\nu)k^2}+f^{\nu b}_{\psi,P}\right)\omega\tau^2\delta^{0}_{\nu,k_1}\delta^{0}_{b,k_2}\,,\nonumber\\
\sigma_\nu=&f^{\nu b}_{\sigma,P}(k\tau)^2\delta^{0}_{\nu,k_1}\delta^{0}_{b,k_2}\,,\nonumber
\end{align}
with,
\begin{align}
f^{\nu b}_{\psi,P}=&\frac{3R_b R_\nu}{64 (15 + 2 R_\nu)^2 (15 + 4 R_\nu) k^4} \left[(975 + 550 R_\nu + 64 R_\nu^2) k^4\right.\nonumber\\\nonumber
&\left. - 15 (135 + 22 R_\nu) (k_1^2 - k_2^2)^2 +  10 k^2 ((105 + 26 R_\nu) k_1^2 + 3 (55 + 6 R_\nu) k_2^2)\right]\,,\\
f^{\nu b}_{v,P}=&-\frac{R_bR_\nu((1200+R_\nu(2R_\nu(65+24R_\nu)-409))k^2-(15+2R_\nu)(7R_\nu-64)(k_1^2-k_2^2))}{32R_\gamma^2 (15+2R_\nu)(15+4R_\nu)k^2}\nonumber\\
f^{\nu b}_{\sigma,P}=&\frac{R_b}{R_c}f^{\nu c}_{\sigma,P}\,.\nonumber
\end{align}

\subsection{Mixture of compensated and neutrino modes}

\begin{align}
\psi^{(2)}=&\frac{R_\nu R_c(25+2R_\nu)}{192 R_\gamma (75+4R_\nu)}(k^2+k_1^2-k_2^2)\omega\tau^3\delta_{\nu,k_1}^{0}\delta_{CI,k_2}^{0}\,,\nonumber\\
\phi^{(2)}=&\frac{R_\nu R_c(25-2R_\nu)}{192 R_\gamma (75+4R_\nu)}(k^2+k_1^2-k_2^2)\omega\tau^3\delta_{\nu,k_1}^{0}\delta_{CI,k_2}^{0}\,,\nonumber\\
\delta_{c}^{(2)}=&-\frac{R_b}{R_c}\delta_{b}^{(2)}=\left(\frac{3R_\nu}{15+4R_\nu}+\frac{3R_\nu(-15+2R_\nu)}{4(15+2R_\nu)(15+4R_\nu)}\omega\tau\right)\delta^{0}_{\nu,k_1}\delta^{0}_{CI,k_2}\,,\nonumber\\
\delta_{\gamma}^{(2)}=&\frac{R_\nu R_c(175-15R_\nu-2R_\nu)}{48 R_\gamma^2(75+4R_\nu)}(k^2+k_1^2-k_2^2)\omega\tau^3\delta_{\nu,k_1}^{0}\delta_{CI,k_2}^{0}\,,\nonumber\\
\delta_{\nu}^{(2)}=&\frac{R_\nu R_c(25+2R_\nu)}{48 R_\gamma (75+4R_\nu)}(k^2+k_1^2-k_2^2)\omega\tau^3\delta_{\nu,k_1}^{0}\delta_{CI,k_2}^{0}\,,\\
v_{\gamma b}^{(2)}=&\frac{3R_\nu R_c(k^2+k_1^2-k_2^2)}{32R_\gamma^2k^2}\omega\tau^2\delta^{0}_{\nu,k_1}\delta^{0}_{CI,k_2}\,,\nonumber\\
v_{\nu}^{(2)}=&-\frac{25 R_\nu R_c}{384 R_\gamma^2}(k^2+k_1^2-k_2^2)\omega\tau^4\delta_{\nu,k_1}^{0}\delta_{CI,k_2}^{0}\,,\nonumber\\
v_{c}^{(2)}=&\frac{R_\nu R_c(25-2R_\nu)}{960 R_\gamma (75+4R_\nu)}(k^2+k_1^2-k_2^2)\omega\tau^4\delta_{\nu,k_1}^{0}\delta_{CI,k_2}^{0}\,.\nonumber
\end{align}

\section{Liouville equation in terms of brightness tensors}\label{boltzapp}

In this appendix we derive the Liouville equation (collisionless Boltzmann equation) for neutrinos at second order in cosmological perturbation theory, in synchronous gauge, and define the different moments of the distribution function and the method to obtain evolution equations for each of them in real space. We follow the notation and conventions of Ref.~\cite{Pettinari:2014vja}, except where indicated.

We are interested here in the evolution of neutrinos after decoupling, i.e., after collisions have become negligible. Then, the Boltzmann equation reduces to the Liouville equation:
\begin{equation}
\frac{df}{d\lambda}=0\,,
\end{equation}
in which $f$ is the distribution function for neutrinos and $\lambda$ is the affine parameter labelling the geodesics followed by the neutrinos.

We begin by defining a tetrad basis, which will make some calculations easier. In general, it is defined such that the metric is the Minkowski metric $\eta_{\ul{a}\ul{b}}$, when evaluated in this basis:
\begin{equation}
g_{\mu\nu}e_{\ul{a}}^\mu e_{\ul{b}}^\nu=\eta_{\ul{a}\ul{b}}\,.
\end{equation}
The components of the basis vectors, $e_{\ul{a}}^\mu$, then represent all the information included in the metric. There are, however, more degrees of freedom in the tetrad components (16) than in the metric (10) and the remaining ones represent 3 Lorentz boosts and 3 rotations. In order to fix the ambiguity in the choice of frame, we set $e_{\ul{0}}^i=0$ and $e^j_{\ul{i}}=e^i_{\ul{j}}$ \cite{Pettinari:2014vja}. In synchronous gauge, the remaining components are given by
\begin{gather}
e_{\ul{0}}^0=\frac{1}{a}\,, \quad\quad e^{\ul{0}}_0=a\,,\\
e_{\ul{0}}^i=0\,, \quad\quad e^{\ul{i}}_0=0\,,\\
e_{\ul{i}}^0=0\,, \quad\quad e^{\ul{0}}_i=0\,,\\
e_{\ul{i}}^j=\frac1a\left[\delta^j_i-C^j_{\ i}+\frac32C^k_{\ i}C^j_{\ k}\right]\,, \quad\quad e^{\ul{j}}_i=a\left[\delta^j_i+C^j_{\ i}-\frac12C^k_{\ i}C^j_{\ k}\right]\,.
\end{gather}
We use this basis to parametrize the 4-momentum vector as 
\begin{equation}
p^{\ul{a}}=(p,p n^i)\,,
\end{equation}
in which we assumed neutrinos to be massless and we defined the magnitude of the momentum, $p$, and the direction of propagation, $n^i$. As with any other vector, the components of the 4-momentum in the coordinate basis are related to those in the tetrad basis via $p^\mu=e^\mu_{\ul{a}}p^{\ul{a}}$. Using this basis, we are now able to write the Liouville equation by expanding the total derivative:
\begin{equation}
\label{boltzeq1}
\frac{\p f}{\p \tau}+\frac{\p f}{\p x^i}\frac{d x^i}{d \tau}+\frac{\p f}{\p p}\frac{d p}{d \tau}+\frac{\p f}{\p n^i}\frac{d n^i}{d \tau}=0\,.
\end{equation}
Using the definition of the 4-momentum we find
\begin{equation}
\frac{d x^i}{d \tau}=\frac{p^i}{p^0}=\left(\delta_j^i-C^i_{\ j}\right)n^j\,.
\end{equation}
From the geodesic equation, we get
\begin{equation}
\frac{1}{p}\frac{d p}{d \tau}=-\left[\Hh \delta_{kl}+C_{kl}'-C^{i}_{\ k}C_{il}'-C^{i}_{\ l}C_{ik}'\right]n^k n^l\,,
\end{equation}
and
\begin{equation}
\frac{d n^i}{d \tau}=-\left(\delta^{ik}-n^in^k\right)\left[C_{kl}'n^l+\psi_{,k}\right]\,.
\end{equation}
We also integrate the Liouville equation, Eq.~\eqref{boltzeq1}, in the momentum magnitude $p$ and rewrite the equation in terms of the brightness fluctuation $\Delta$, defined by
\begin{equation}
\Delta(\tau,\vec x,\vec n)=\frac{\int{{\rm d}p\, p^3 (f(\tau,\vec x,p,\vec n)-\bar f(\tau,p))}}{\int{{\rm d}p\, p^3 \bar f(\tau,p)}}\,,
\end{equation}
where $\bar f$ is the background neutrino distribution function. The momentum integrated Liouville equation in synchronous gauge is therefore given by
\begin{align}
\label{Boltzint}
\Delta'+\p_i\Delta\left(\delta^i_j-C^i_{\ j}\right)n^j+4(1+\Delta) n^k n^l \left(C_{kl}'-C^{i}_{\ k}C_{il}'-C^{i}_{\ l}C_{ik}'\right)\\
-\frac{\p \Delta}{\p n^i}\left(\delta^{ik}-n^in^k\right)\left[C_{kl}'n^l+\psi_{,k}\right]=0\nonumber\, .
\end{align}
This is a partial differential equation in $\tau$, $x^i$ and $n^i$. In order to simplify its solution, we integrate out its angular dependence. This procedure will generate a set of equations, each of which obtained by a different weight in the angular integral. The most common way to do this is to project the Liouville equation in multipole space as is done in Ref.~\cite{Pettinari:2014vja}. In this work, however, we choose to do something slightly different, and introduce instead a tensorial projection. The main difference is that, instead of using spherical harmonics as weights, we use the direction vector $n^i$ in different powers. To clarify, the projectors being used here to extract each equation are given by
\begin{equation}
\label{projn}
\mathcal{P}_N^{i_1\cdots i_N}=\int{\frac{{\rm d}\Omega}{4\pi} n^{i_1}\cdots n^{i_N}}\,,
\end{equation}
in which the integral is over all possible angular directions and the measure, $\frac{{\rm d}\Omega}{4\pi}$, is such that $\int{\frac{{\rm d}\Omega}{4\pi}}=1$. The application of these projectors to the brightness fluctuation generates the brightness tensors, shown here up to rank 3
\begin{align}
\Delta_0=\mathcal{P}_0 [\Delta]=&\int{\frac{{\rm d}\Omega}{4\pi} \Delta(\tau,\vec x,\vec n) }\,,\\
\Delta^{i}=\mathcal{P}_1^i [\Delta]=&\int{\frac{{\rm d}\Omega}{4\pi} n^i \Delta(\tau,\vec x,\vec n) }\,,\\
\Delta^{ij}=\mathcal{P}_2^{ij} [\Delta]=&\int{\frac{{\rm d}\Omega}{4\pi} n^i n^j \Delta(\tau,\vec x,\vec n) }\,,\\
\Delta^{ijk}=\mathcal{P}_3^{ijk} [\Delta]=&\int{\frac{{\rm d}\Omega}{4\pi} n^i n^j n^k \Delta(\tau,\vec x,\vec n) }\,.
\end{align}
Note that these tensors appear to describe more degrees of freedom than the usual multipoles. For example, $\Delta^{ij}$ is a symmetric 3-tensor, thus having in total 6 degrees of freedom, while the $\ell=2$ multipoles only represent $2\ell+1=5$ degrees of freedom. This discrepancy can be understood by noticing that the brightness tensors are related amongst each other. The extra d.o.f. in this example is actually in the trace of $\Delta^{ij}$, which is obviously equal to $\Delta_0$, since $n_in^i=1$. Therefore, it is the traceless part of each of these tensors that includes the same information as the usual multipoles. For that reason, it is useful to also define traceless brightness tensors:
\begin{align}
\Delta_T^{ij}=&\Delta^{ij}-\frac13\delta^{ij}\Delta_0\,,\\
\Delta_T^{ijk}=&\Delta^{ijk}-\frac{3}{5}\delta^{(ij}\Delta^{k)}\,,\\
\Delta_T^{ijkl}=&\Delta^{ijkl}-\frac{6}{7}\delta^{(ij}\Delta_T^{kl)}-\frac{1}{5}\delta^{(ij}\delta^{kl)}\Delta_0\,,\\
\Delta_T^{ijklm}=&\Delta^{ijklm}-\frac{10}{9}\delta^{(ij}\Delta_T^{klm)}-\frac{3}{7}\delta^{(ij}\delta^{kl}\Delta^{m)}\,.
\end{align}
These are the tensors for which we are interested in finding evolution equations. In order to do that, we simply project the momentum integrated Liouville equation, Eq.~\eqref{Boltzint}, with the projectors $\mathcal{P}_N$ defined in Eq.~\eqref{projn}. For each value of $N$ this procedure will result in an evolution equation for the corresponding brightness tensor of rank $N$. Obtaining the equations for the traceless tensors is straightforward by subtracting the corresponding trace equation. Eq.~\eqref{Boltz}, was obtained through this method and using it for $N=0$ and $N=1$ would reproduce the conservation of the stress-energy tensor for neutrinos. This can be seen by noting that the relation between the stress-energy tensor and the distribution function is given by
\begin{equation}
T^{\ul{a}}_{\ \ \ul{b}}=\int\text{d}^3p \frac{p^{\ul{a}}p_{\ul{b}}}{p} f\,.
\end{equation}
This relation was used to derive Eqs.~\eqref{d0}--\eqref{d2}, describing the lowest rank brightness tensors in terms of the perturbed stress-energy tensor components in the coordinate basis. We now show the explicit version of that relation, specialising only to the scalar parts of brightness tensors:
\begin{equation}
\Delta_0=\delta_\nu+\frac43v_{\nu,i}v_{\nu}^{,i}\,,
\end{equation}
\begin{equation}
\n^2\Delta_1=\frac43\n^2v_{\nu}+\p_i\left[\left(\frac43(\delta_\nu-\psi)\delta^i_j+\frac43 E^{,i}_{,j}+\frac{1}{\rho_\nu}(\Pi_{\nu,j}^{,i}-\frac13\delta^i_j\n^2\Pi_\nu) \right)v_\nu^{,j}\right]\,,
\end{equation}
\begin{align}
\n^2\n^2\Delta_2=&-2\n^2\sigma_{\nu}+\p_i\p^j\left[2v_{\nu,j}v_{\nu}^{,i}-\frac23v_{\nu,k}v_{\nu}^{,k}\delta^i_j+\frac6{\rho_\nu}\psi\left(\Pi_{\nu,j}^{,i}-\frac13\delta^i_j\n^2\Pi_\nu\right)\right.\\
&\left.-\frac{1}{\rho_\nu}\left(\frac32\Pi_{\nu,jk}E^{,ki}+\frac32\Pi_{\nu,k}^{,i}E_{,j}^{,k}-\n^2\Pi_{\nu}E_{,j}^{,i}+\left(\frac13\n^2\Pi_{\nu}\n^2E-\Pi_{\nu,kl}E^{,kl}\right)\delta^i_j\right)\right]\,.\nonumber
\end{align}
The scalar variables denoted above by $\Delta_N$ are the scalar parts of the brightness tensors of rank $N$. They are obtained by performing the scalar-vector-tensor decomposition of those tensors. We now describe that decomposition for the brightness tensors up to rank 3, which are used in the main text. For the rank 1 and 2 tensors, we use the same decomposition as for the velocity and anisotropic stress, respectively:
\begin{equation}
\Delta^{i}=\Delta_1^{,i}+\Delta_{1v}^i\,,
\end{equation}
\begin{equation}
\Delta_T^{ij}=\Delta_{2}^{,ij}-\frac{1}{3}\delta^{ij}\n^2\Delta_2+\Delta_{2v}^{(i,j)}+\Delta_{2t}^{ij}\,.
\end{equation}
The labels $v$ and $t$ denote the transverse vector and transverse and traceless tensor parts. As for the rank 3 tensor, there are, in total, 7 degrees of freedom split into one scalar, one vector, one rank 2 tensor and one rank 3 tensor. They are defined via
\begin{equation}
\Delta_T^{ijk}=\Delta_3^{,ijk}-\frac35\delta^{(ij}\n^2\Delta_3^{,k)}+\Delta_{3v}^{(i,jk)}-\frac15\delta^{(ij}\n^2\Delta_{3v}^{k)}+\Delta_{3t}^{(ij,k)}+\Delta_{3T}^{ijk}\,.
\end{equation}
The functional form of this splitting was derived by writing the most general expression including all the degrees of freedom and then applying the traceless and symmetric conditions in all indices to find the appropriate coefficient values.
\\\\
Finally, we address the issues related to the projection of the distribution function in terms of brightness tensors and its relation to the more common projection in multipole space. We aim here to find a relation between the two so that our results may be translatable to that formalism and vice versa.

We begin with the definition of the multipole projection of the brightness fluctuation, $\Delta$:
\begin{equation}
\Delta_{\ell m}=i^\ell\sqrt{\frac{2\ell+1}{4\pi}}\int{{\rm d}\Omega Y_{\ell,m}^* \Delta}\,,\label{dlm}
\end{equation}
in which $Y_{\ell,m}^*$ is the complex conjugate of the spherical harmonic $Y_{\ell,m}$. To explicitly demonstrate the connection to our formalism, we apply the appropriate transformations to our definitions to obtain a new version in terms of the multipole decomposition. Since we are only interested in the scalar variables, we begin by taking $N$ spatial derivatives in a rank $N$ brightness tensor to extract its scalar part. In Fourier space, this is equivalent to contracting those tensors $N$ times with $i k^i$. For the example case of rank 1, this results in
\begin{equation}
i k_i\Delta^{i}=i \int{\frac{{\rm d}\Omega}{4\pi} k_i n^i \Delta }\,.\label{kdelta1}
\end{equation}
By definition of the scalar product we have $k_i n^i=k \cos \theta$ and it can be easily verified that $\cos\theta\propto Y_{10}$. After some algebra, we can transform Eq.~\eqref{kdelta1} into the form of Eq.~\eqref{dlm} for $\ell=1$ and $m=0$, thus showing that
\begin{equation}
\Delta_{10}=-3k\Delta_1\,.
\end{equation}
Generalising this procedure for higher $\ell$ and correspondingly higher rank tensors is conceptually straightforward and it can be shown that the general relation is simply given by
\begin{align}
\Delta_{\ell0}=(-1)^\ell(2\ell+1)k^\ell\Delta_\ell\,.
\end{align}
With this simple relation, the interested reader can translate all our results into the multipole formalism with ease.

\bibliographystyle{JHEPmodplain}
\bibliography{isocurvatures}

\end{document}